\documentclass[oneauthor,pdftex,book]{Definitions/mdpi}
\usepackage[utf8]{inputenc}
\usepackage{array}
\usepackage{enumitem}
\usepackage{amsmath}
\usepackage{amsthm}
\usepackage{graphicx}
\usepackage{wasysym}

\usepackage{array}
\newcolumntype{x}[1]{>{\centering\let\newline\\\arraybackslash\hspace{0pt}}p{#1}}

\makeatletter
\abstract{\MyAbstract}

\providecommand{\tabularnewline}{\\}

\theoremstyle{plain}
\newtheorem{thm}{\protect\theoremname}
\theoremstyle{plain}
\newtheorem{lem}[thm]{\protect\lemmaname}
\theoremstyle{plain}
\newtheorem{cor}[thm]{\protect\corollaryname}

\usepackage{stmaryrd}

\firstpage{1} 
 
\pubvolume{1}
\issuenum{1}
\articlenumber{0}
\pubyear{2022}
\copyrightyear{2022}
\externaleditor{Academic Editor: Eckehard Olbrich} 
\datereceived{4 January 2022} 
\dateaccepted{23 February 2022} 
\datepublished{} 
\hreflink{https://doi.org/} 
\makeatletter 

\usepackage[capitalise]{cleveref}
\crefname{thm}{Theorem}{Theorems}
\crefname{lem}{Lemma}{Lemmas}
\crefname{prop}{Proposition}{Propositions}
\crefname{defn}{Definition}{Definitions}
\crefname{rem}{Remark}{Remarks}

\Title{A Novel Approach to the Partial Information Decomposition}
\TitleCitation{A Novel Approach to the Partial Information Decomposition}
\abstract{\textls[-5]{We consider the ``partial information decomposition'' (PID) problem, 
which aims to decompose the information that a set of source random variables provide 
about a target random variable into separate redundant, synergistic, union, and unique components. 
In the first part of this paper, we propose a general framework for constructing a multivariate PID. 
Our framework is defined in terms of a formal analogy with intersection and union from set theory, 
along with an ordering relation which specifies when one information source is more informative than another. 
Our definitions are  algebraically and axiomatically motivated, and can be generalized to
 domains beyond Shannon information theory (such as algorithmic information theory and quantum information theory). 
In the second part of this paper, we use our general framework to define a PID in terms of the well-known 
Blackwell order, 
which has a fundamental operational interpretation. 
We demonstrate our approach on numerous examples and show that it overcomes many drawbacks associated with previous proposals.}}
\Author{
Artemy Kolchinsky 
}

\AuthorNames{Artemy Kolchinsky}

\AuthorCitation{Kolchinsky, A.}
\address[1]{%
Santa Fe Institute, Santa Fe, NM {87501, USA}; artemyk@gmail.com}

\ifdef{\articlenumber}{

\newcommand*\patchAmsMathEnvironmentForLineno[1]{%
  \expandafter\let\csname old#1\expandafter\endcsname\csname #1\endcsname
  \expandafter\let\csname oldend#1\expandafter\endcsname\csname end#1\endcsname
  \renewenvironment{#1}%
     {\linenomath\csname old#1\endcsname}%
     {\csname oldend#1\endcsname\endlinenomath}}%
\newcommand*\patchBothAmsMathEnvironmentsForLineno[1]{%
  \patchAmsMathEnvironmentForLineno{#1}%
  \patchAmsMathEnvironmentForLineno{#1*}}%
\AtBeginDocument{%
\patchBothAmsMathEnvironmentsForLineno{equation}%
\patchBothAmsMathEnvironmentsForLineno{align}%
\patchBothAmsMathEnvironmentsForLineno{flalign}%
\patchBothAmsMathEnvironmentsForLineno{alignat}%
\patchBothAmsMathEnvironmentsForLineno{gather}%
\patchBothAmsMathEnvironmentsForLineno{multline}%
}

}

\makeatother

\providecommand{\corollaryname}{Corollary}
\providecommand{\lemmaname}{Lemma}
\providecommand{\theoremname}{Theorem}

\begin{document}

\section{Introduction}

\textls[-14]{Understanding how information is distributed in multivariate systems
is an important problem in many scientific fields. In the context
of neuroscience, for example, one may wish to understand how information
about an external stimulus is encoded in the activity of different
brain regions. In computer science, one might wish to understand how
the output of a logic gate reflects the information present in different
inputs to that gate. Numerous other examples abound in biology, physics,
machine learning, cryptography, and other fields \citep{schneidman_synergy_2003,daniels_quantifying_2016,tax2017partial,amjad2018understanding,lizier2018information,wibral_partial_2017,timme_synergy_2014,chan_multivariate_2015,rosas2020reconciling,cang2020inferring}.}

Formally, suppose that we are provided with a random variable $Y$
which we call the ``target'', as well as a set of $n$ random variables
$X_{1},\dots,X_{n}$ which we call the ``sources''. The \emph{partial
information decomposition} (PID), first proposed by Williams and Beer
in 2010 \citep{williams2010nonnegative}, aims to quantify how information
about the target is distributed among the different sources. In particular,
the PID seeks to decompose the mutual information provided jointly
by all sources into a set of nonnegative terms, such as \emph{redundancy
}(information present in each individual source), \emph{synergy} (information
only provided by the sources jointly, not individually), \emph{union
information} (information provided by at least one individual source),
and \emph{unique information} (information provided by only one individual
source).

As discussed in detail below, the PID is inspired by an analogy between
information theory and set theory. In this analogy, the information
that the sources provide about the target are imagined as sets, while
PID terms such as redundancy, union information, and synergy are imagined
as the sizes of intersections, unions, and complements. While the
analogy between information-theoretic and set-theoretic quantities
is suggestive, it does not specify how to actually define the PID.
Moreover, it has also been shown that existing measures from information
theory (such as mutual information and conditional mutual information)
cannot be used directly to construct the PID, since these measures
conflate contributions from different terms like synergy and redundancy
\citep{williams2010nonnegative,williams_information_2011}. In response,
many proposals for how to define PID terms have been advanced \citep{bertschinger2014quantifying,quax2017quantifying,james2018unique,griffith2014intersection,griffith2014quantifying,griffith_quantifying_2015,harder_bivariate_2013,ince_measuring_2017,finn_pointwise_2018,lizier2018information}.
However, existing proposals suffer from various drawbacks, such as
behaving counterintuitively on simple examples, being limited to
only two sources, or lacking a clear operational interpretation. Today
there is no generally agreed-upon way of defining the PID.

In this paper, we propose a new and principled approach to the PID
which addresses these drawbacks. Our approach can handle any number
of sources and can be justified in algebraic, axiomatic, and operational
terms. We present our approach in two parts.

In part I (\cref{sec:framework}), we propose a general framework for
defining the PID. Our framework does not prescribe specific definitions,
but instead shows how an information-theoretic decomposition can be
grounded in a formal analogy with set theory. Specifically, we consider
the definitions of ``set intersection'' and ``set union'' in set
theory: the intersection of sets $S_{1},S_{2},\dots$ is the largest
set that is contained in all of the $S_{i}$, while the union of sets
$S_{1},S_{2},\dots$ is the smallest set that contains all of the
$S_{i}$. As we show, these set-theoretic definitions can be mapped
into information-theoretic terms by treating ``sets'' as random
variables, ``set size'' as mutual information between a random variable
and the target $Y$, and ``set inclusion'' as some externally specified
ordering relation $\sqsubset$, which specifies when one random variable
is more informative than another. Using this mapping, we define information-theoretic
redundancy and union information in the same way that the sizes of
intersections and unions are defined in set theory (other PID terms,
such as synergy and unique information, can be computed in a straightforward
way from redundancy and union information). Moreover, while our approach
is motivated by set-theoretic intuitions, as we show in \cref{subsec:Axiomatic-derivation},
it can also be derived from an alternative axiomatic foundation. We
finish part I by reviewing relevant prior work in information theory
and the PID literature. We also discuss how our framework can be generalized
beyond the standard setting of the PID and even beyond Shannon information
theory, to domains like algorithmic information theory and quantum
information theory.

\textls[-25]{One unusual aspect of our framework is that it provides independent
definitions of union information and redundancy. Most prior work on
the PID has focused exclusively on the definition of redundancy, because
it assumed that union information can be determined from redundancy
using the so-called ``inclusion-exclusion principle''. In \cref{sec:iep-violation},
we argue that the inclusion-exclusion principle should not be expected
to hold in the context of the PID.}

Part I provides a general framework. Concrete definitions of the PID
can be derived from this general framework by choosing a specific
``more informative'' ordering relation $\sqsubset$. In fact, the
study of ordering relations between information sources has a long
history in statistics and information theory \citep{shannon_lattice_1953,shannon_note_1958,cohen1998comparisons,le1964sufficiency,korner1977comparison,torgersen1991comparison}.
One particularly important relation is the so-called ``Blackwell
order'' \citep{blackwell_equivalent_1953,bertschinger2014quantifying},
which has a fundamental operational interpretation in terms of utility
maximization in decision theory.

\textls[-33]{In part II of this paper (\cref{sec:partwo}), we combine the general
framework developed in part I with the Blackwell order. This gives
rise to concrete definitions of redundancy and union information.
We show that our measures behave intuitively and have simple operational
interpretations in terms of decision theory. Interestingly, while
our measure of redundancy is novel, our measure of union information
has previously appeared in the literature under a different guise
\citep{bertschinger2014quantifying,griffith2014quantifying}.}

In \cref{sec:Relation-to-existing}, we compare our redundancy measure
to previous proposals, and illustrate it with various bivariate and
multivariate examples. We finish the paper with a discussion and proposals
for future work in \cref{sec:Discussion}.

We introduce some necessary notation and preliminaries in the next
section. In addition, we provide background regarding the PID in \cref{sec:Background}.
All proofs, as well as some additional results, are found in the appendix.

\section{Notation and Preliminaries}

We use uppercase letters ($Y,X,Q,\dots$) to indicate random variables
over some underlying probability space. We use lowercase letters ($y,x,q,\dots$)
to indicate specific outcomes of random variables, and calligraphic
letters ($\mathrm{\mathcal{Y}},\mathcal{X},\mathcal{Q}\dots$) to
indicate sets of outcomes. We often index random variables with a
subscript, e.g., the random variable $X_{i}$ with outcomes $x_{i}\in\mathcal{X}_{i}$
(so $x_{i}$ does not refer to the $i^{\mathrm{th}}$ outcome of random
variable $X$, but rather to some generic outcome of random variable
$X_{i}$). We use notation like $A-B-C$ to indicate that $A$ is
conditionally independent of $C$ given $B$. Except where otherwise
noted, we assume that all random variables have a finite number of
outcomes. 

We use notation like $P_{X}(x)$ to indicate the probability distribution
associated with random variable $X$, $P_{XY}(x,y)$ to indicate the
joint probability distribution associated with random variables $X$
and $Y$, and $P_{X\vert Y}(x\vert y)$ to indicate the conditional
probability distribution of $X$ given $Y$. Given two random variables
$X$ and $Y$ with outcome sets $\mathcal{X}$ and $\mathrm{\mathcal{Y}}$,
we use notation like $\kappa_{X\vert Y}(x\vert y)$ to indicate some
stochastic \emph{channel} of outputs $x\in\mathcal{X}$ given inputs
$y\in\mathrm{\mathcal{Y}}$. In general, a channel $\kappa_{X\vert Y}$
specifies some arbitrary conditional distribution of $X$ given $Y$,
which can be different from $P_{X\vert Y}$, the actual conditional
distribution of $X$ given $Y$ (as determined by the underlying probability
space).

As described above, we consider the information that a set of ``source''
random variables $X_{1},\dots,X_{n}$ provide a ``target'' random
variable $Y$. Without loss of generality, we assume that the marginal
distributions $P_{Y}$ and $P_{X_{i}}$ for all $i$ have full support
(if they do not, one can restrict $\mathbb{\mathrm{\mathcal{Y}}}$
and/or $\mathcal{X}_{i}$ to outcomes that have strictly positive
probability). 

{Finally, note that despite our use of the terms ``source'' and
``target'', we do not assume any causal directionality between the
sources and target (see also discussion in \citep{james2018perspective}).
For example, in neuroscience, $Y$ might be an external stimulus which
causes the activity of brain regions $X_{1},\dots,X_{n}$, while in
computer science $Y$ might represent the output of a logic gate caused
by inputs $X_{1},\dots,X_{n}$ (so the causal direction is reversed).
In yet other contexts, there could be other causal relationships among
$X_{1},\dots,X_{n}$ and $Y$, or they might not be causally related
at all.}

\section{Background on the Partial Information Decomposition (PID)}

\label{sec:Background}\label{sec:PID}

Given a set of sources $X_{1},\dots,X_{n}$ and a target $Y$, the
PID aims to decompose  $I(Y;X_{1},\dots,X_{n})$, the total mutual
information provided by all sources about the target, into a set of
nonnegative terms such as \citep{williams2010nonnegative,williams_information_2011}:

\vspace{10pt}
\noindent \emph{Redundancy} $I_{\cap}(X_{1};\dots;X_{n}\!\shortrightarrow\!Y)$,
the information present in each individual source. Redundancy can
be considered as the intersection of the information provided by different
sources and is sometimes called ``intersection information'' in
the literature \citep{griffith2014intersection,griffith_quantifying_2015}.

\vspace{5pt}
\noindent \emph{Union information} $I_{\cup}(X_{1};\dots;X_{n}\!\shortrightarrow\!Y)$,
the information provided by at least one individual source \citep{williams_information_2011,griffith2014quantifying}.

\vspace{5pt}
\noindent \emph{Synergy} $S(X_{1};\dots;X_{n}\!\shortrightarrow\!Y)$, the information
found in the joint outcome of all sources, but not in any of their
individual outcomes. Synergy is defined as \citep{griffith2014quantifying}
\begin{align}
S(X_{1};\dots;X_{n}\!\shortrightarrow\!Y) & =I(Y;X_{1},\dots,X_{n})-I_{\cup}(X_{1};\dots;X_{n}\!\shortrightarrow\!Y).\label{eq:synDefM}
\end{align}

\vspace{10pt}
\noindent \emph{Unique information} in source $X_{i}$, $U(X_{i}\!\shortrightarrow\!Y\vert X_{1};\dots;X_{n})$,
the non-redundant information in each particular source. Unique information
is defined as
\begin{align}
U(X_{i}\!\shortrightarrow\!Y\vert X_{1};\dots;X_{n}) & =I(Y;X_{i})-I_{\cap}(X_{1};\dots;X_{n}\!\shortrightarrow\!Y).\label{eq:unqdef}
\end{align}

In addition to the above terms, one can also define \emph{excluded
information},
\begin{equation}
E(X_{i}\!\shortrightarrow\!Y\vert X_{1};\dots;X_{n})=I_{\cup}(X_{1};\dots;X_{n}\!\shortrightarrow\!Y)-I(Y;X_{i}),\label{eq:excdef}
\end{equation}
as the information in the union of the sources which is not in a particular
source $X_{i}$. To our knowledge, excluded information has not been
previously considered in the PID literature, although it is the natural
``dual'' of unique information as defined in \cref{eq:unqdef}.

\begin{figure}[t!]
\begin{centering}
\includegraphics[width=0.8\columnwidth]{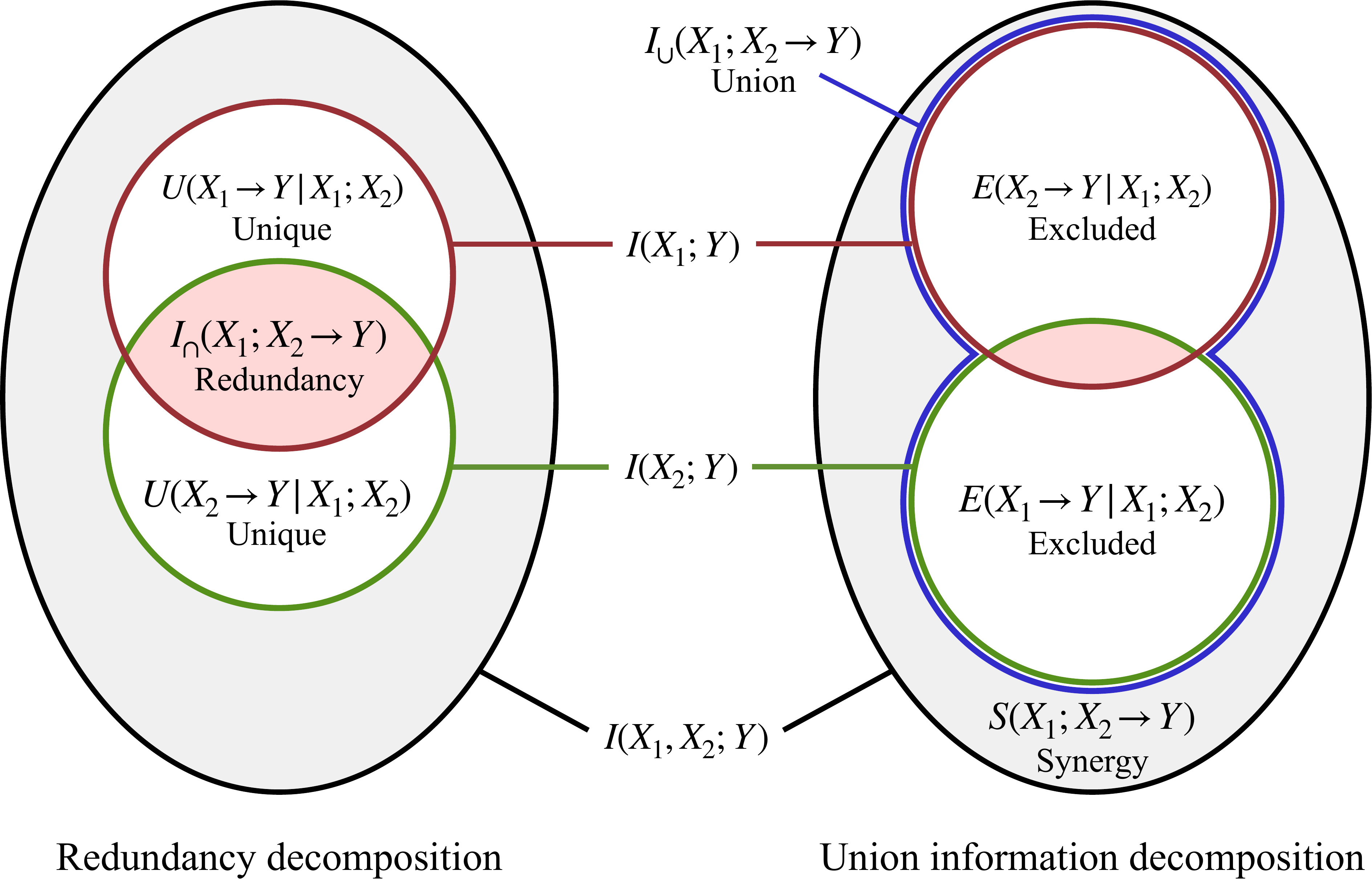}
\par\end{centering}
\caption{\label{fig:venn}Partial information decomposition of the information
provided by two sources about a target. On the left, we show the decomposition
induced by redundancy $I_{\cap}$, which leads to measures of unique
information $U$. On the right, we show the decomposition induced
by union information $I_{\cup}$, which leads to measures of synergy
$S$ and excluded information $E$.}
\end{figure}

Given the definitions above, once a measure of redundancy $I_{\cap}$
is chosen, unique information is determined by \cref{eq:unqdef}.
Similarly, once a measure of union information $I_{\cup}$ is chosen,
synergy and excluded information are determined by \cref{eq:synDefM,eq:excdef}.
In \cref{fig:venn}, we illustrate the relationships between these
different PID terms for the simple case of two sources, $X_{1}$ and
$X_{2}$. We show two different decompositions of the information
provided by the sources jointly, $I(X_{1},X_{2};Y)$, and individually,
$I(X_{1};Y)$ and $I(X_{2};Y)$. The diagram on the left shows the
decomposition defined in terms of redundancy $I_{\cap}$, while the
diagram on the right shows the decomposition defined in terms of union
information $I_{\cup}$.

When more than two sources are present, the PID can be used to define
additional terms, beyond the ones shown in \cref{fig:venn}. For example,
for three sources, one can define redundancy terms like $I_{\cap}(X_{1},X_{2},X_{3}\!\shortrightarrow\!Y)$
(representing the information found in all individual sources) as
well as redundancy terms like $I_{\cap}((X_{1},X_{2}),(X_{1},X_{3}),(X_{2},X_{3})\!\shortrightarrow\!Y)$
(representing the information found in all pairs of sources), and
similarly for union information.

The idea that redundancy and union information lead to two different
information decompositions is rarely discussed in the literature.
In fact, the very concept of union information is rarely discussed
in the literature explicitly (although it often appears in an implicit
form via measures of synergy, since synergy is related to union information
through \cref{eq:synDefM}). As we discuss below in \cref{sec:iep-violation},
the reason for this omission is that most existing work assumes (whether
implicitly or explicitly) that redundancy and union information are
not independent measures, but are instead related via the so-called
``inclusion-exclusion principle''. If the inclusion-exclusion principle
is assumed to hold, then the distinction between the two decompositions
disappears. We discuss this issue in greater detail below, where we
also argue that the inclusion-exclusion principle should not be expected
to hold in the context of the PID. 

We have not yet described how the redundancy and union information
measures $I_{\cap}$ and $I_{\cup}$ are defined. In fact, this remains
an open research question in the field (and one which this paper will
address). When they first introduced the idea of the PID, Williams
and Beer proposed a set of intuitive axioms that any measure of redundancy
should satisfy \citep{williams2010nonnegative,williams_information_2011},
which we summarize in \cref{app:axioms}. In later work, Griffith and
Koch \citep{griffith2014quantifying} proposed a similar set of axioms
that union information should satisfy, which are also summarized in
\cref{app:axioms}. However, these axioms do not uniquely identify
a particular measure of redundancy or union information.

Williams and Beer also proposed a particular redundancy measure which
satisfies their axioms, which we refer to as $I_{\cap}^{\mathrm{WB}}$
\citep{williams2010nonnegative,williams_information_2011}. Unfortunately,
$I_{\cap}^{\mathrm{WB}}$ has been shown to behave counterintuitively
in some simple cases \citep{harder_bivariate_2013,ince_measuring_2017}.
For example, consider the so-called ``COPY gate'', where there are
two sources $X_{1}$ and $X_{2}$ and the target is a copy of their
joint outcomes, $Y=(X_{1},X_{2})$. If $X_{1}$ and $X_{2}$ are statistically
independent, $I(X_{1};X_{2})=0$, then intuition suggests that the
two sources provide independent information about $Y$ and therefore
that redundancy should be 0. In general, however, $I_{\cap}^{\mathrm{WB}}(X_{1};X_{2}\!\shortrightarrow\!Y)$
does not vanish in this case. To avoid this issue, Ince \citep{ince_measuring_2017}
proposed that any valid redundancy measure should obey the following
property:
\begin{equation}
\text{If }I(X_{1};X_{2})=0\text{, then }I_{\cap}(X_{1};X_{2}\!\shortrightarrow\!(X_{1},X_{2}))=0,\label{eq:iiprop}
\end{equation}
which is called the \emph{Independent identity property}.

In recent years, many other redundancy measures have been proposed
\citep{ince_measuring_2017,bertschinger2014quantifying,harder_bivariate_2013,james2018unique,finn_pointwise_2018,griffith2014intersection,griffith_quantifying_2015}.
However, while some of these proposals satisfy the Independent identity
property, they suffer various other drawbacks, such as exhibiting
other types of counterintuitive behavior, being limited to two sources,
and/or lacking a clear operational motivation. We discuss some of
these previously proposed measures in \cref{subsec:Relation-to-prior,subsec:Blackwell-union-information-priorwork,sec:Relation-to-existing}.

Unlike redundancy, to our knowledge only two measures of union information
have been advanced. The first one appeared in the original work on
the PID \citep{williams_information_2011}, and was derived from $I_{\cap}^{\mathrm{WB}}$
using the inclusion-exclusion principle. The second one appeared more
recently \citep{griffith2014quantifying,bertschinger2014quantifying} 
and is discussed in \cref{subsec:Blackwell-union-information-priorwork}
below.

\section{Part I: Redundancy and Union Information from an Ordering Relation}

\label{sec:framework}

\subsection{Introduction}

\label{subsec:RedundancyDef}

As mentioned above, PID is motivated by an informal analogy with set
theory \citep{williams_information_2011}. In particular, redundancy
is interpreted analogously to the size of the intersection of the
sources $X_{1},\dots,X_{n}$, while union information is interpreted
analogously to the size of their union. 

We propose to define the PID by making this analogy formal, and in
particular by going back to the algebraic definitions of intersection
and union in set theory. In pursuing this direction, we build on a
line of previous work in information theory and PID, which we discuss
in \cref{subsec:Relation-to-prior}.

Recall that in set theory, the intersection of sets $S_{1},\dots,S_{n}\subseteq U$
(where $U$ is some universal set) is the largest set that is contained
in all $S_{i}$ (Section 7.2, \cite{whitelaw_introduction_1988}).
This means that the size of the intersection can be written as
\begin{align}
\Big|\bigcap_{i}S_{i}\Big| & =\sup_{T\subseteq U}\;\left|T\right|\;\;\text{such that}\;\forall i\;T\subseteq S_{i},\label{eq:int0}
\end{align}
Similarly, the union of sets $S_{1},\dots,S_{n}\subseteq U$ is the
smallest set that contains all $S_{i}$ (Section 7.2, \cite{whitelaw_introduction_1988}),
so the size of the union can be written as
\begin{align}
\Big|\bigcup_{i}S_{i}\Big| & =\inf_{T\subseteq U}\;\left|T\right|\;\;\text{ \text{such that} }\;\forall i\;S_{i}\subseteq T.\label{eq:union0-1}
\end{align}
\cref{eq:int0,eq:union0-1} are useful because they express the size
of the intersection and union via an optimization over simpler terms
(the size of individual sets, $\vert T\vert$, and the subset inclusion
relation, $\subseteq$).

We translate these definitions to the information-theoretic setting
of the PID. We take the analogue of a ``set'' to be some random
variable $A$ that provides information about the target $Y$, and
the analogue of ``set size'' to be the mutual information $I(A;Y)$.
In addition, we assume that there is some ordering relation $\sqsubset$
between random variables analogous to set inclusion $\subseteq$.
Given such a relation, the expression $A\sqsubset B$ means that random
variable $B$ is ``more informative'' than $A$, in the sense that
the information that $A$ provides about $Y$ is contained within
the information that $B$ provides about $Y$. 

At this point, we leave the ordering relation $\sqsubset$ unspecified.
In general, we believe that the choice of $\sqsubset$ will not be
determined from purely information-theoretic considerations, but may
instead depend on the operational setting and scientific domain in
which the PID is applied. At the same time, there has been a great
deal of research on ordering relations in statistics and information
theory. In part II of this paper, \cref{sec:partwo}, we will combine
our general framework with a particular ordering relation, the so-called
``Blackwell order'', which has a fundamental interpretation in terms
of decision theory. 

We now provide formal definitions of redundancy and union information,
relative to the choice of ordering relation $\sqsubset$. In analogy
to \cref{eq:int0}, we define redundancy as
\begin{align}
I_{\cap}(X_{1};\dots;X_{n}\!\shortrightarrow\!Y) & :=\sup_{Q}I(Q;Y)\;\;\mathrm{\text{such that}}\;\forall i\;Q\sqsubset X_{i}\label{eq:IcapGen}
\end{align}
where the maximization is over all random variables with a finite
number of outcomes. Thus, redundancy $I_{\cap}$ is the maximum information
about $Y$ in any random variable that is less informative than all
of the sources. In analogy with \cref{eq:union0-1}, we define union
information as 
\begin{align}
I_{\cup}(X_{1};\dots;X_{n} & \!\shortrightarrow\!Y):=\inf_{Q}I(Q;Y)\;\;\mathrm{\text{such that}}\;\forall i\;X_{i}\sqsubset Q\label{eq:IcupGen}
\end{align}
Thus, union information $I_{\cup}$ is the minimum information about
$Y$ in any random variable that is more informative than all of the
sources. Given these definitions, other elements of the PID (such
as unique information, synergy, and excluded information) can be defined
using the expressions found in \cref{sec:Background}. Note that $I_{\cap}$
and $I_{\cup}$ depend the choice of ordering relation $\sqsubset$,
although for convenience we leave this dependence implicit in our
notation.

One of the attractive aspects of our definitions is that they do not
simply quantify the amount of redundancy and union information, but
also specify the ``content'' of that redundant and union information.
In particular, the random variable $Q$ that achieves the optimum
in \cref{eq:IcapGen} specifies the content of the redundant information
via the joint distribution $P_{YQ}$. Similarly, the random variable
$Q$ which achieves the optimum in \cref{eq:IcupGen} specifies the
content of the union information via the joint distribution $P_{YQ}$.
Note that these optimizing $Q$ may not be unique, reflecting
the fact that there may be different ways to represent the redundancy
or union information. (Note also that the supremum or infinitum may not
be achieved in \cref{eq:IcapGen,eq:IcupGen}, in which case one can
consider $Q$ that achieve the optimal values to any desired precision
$\epsilon>0$.)

So far we have not made any assumptions about the ordering relation
$\sqsubset$. However, we can derive some useful bounds by introducing
three weak assumptions:
\begin{enumerate}
\item[I.] Monotonicity of mutual information: $A\sqsubset B\implies I(A;Y)\le I(B;Y)$
(less informative sources have less mutual information).
\item[II.] Reflexivity: $A\sqsubset A$ for all $A$ (each source is at least
as informative as itself).
\item[III.] For all sources $X_{i}$, $O\sqsubset X_{i}\sqsubset(X_{1},\dots,X_{n})$,
where $O$ indicates a constant random variable with a single outcome
and $(X_{1},\dots,X_{n})$ indicates all sources considered jointly
(each source is more informative than a trivial source and less informative
than all sources jointly).
\end{enumerate}
Assumptions I and II imply that the redundancy and union information
of a single source are equal to the mutual information in that source:
\[
I_{\cap}(X_{1}\!\shortrightarrow\!Y)=I_{\cup}(X_{1}\!\shortrightarrow\!Y)=I(X_{1};Y).
\]
Assumptions I and III imply the following bounds on redundancy and
union information:
\begin{align}
0 & \le I_{\cap}(X_{1};\dots;X_{n}\!\shortrightarrow\!Y)\le\min_{i}I(Y;X_{i})\,.\label{eq:bound1}\\
\max_{i}I(Y;X_{i}) & \le I_{\cup}(X_{1};\dots;X_{n}\!\shortrightarrow\!Y)\le I(Y;X_{1},\dots,X_{n}).\label{eq:boundunion-1}
\end{align}
\cref{eq:bound1} in turn implies that the unique information in each
source $X_{i}$, as defined in \cref{eq:unqdef}, is bounded between
$0$ and $I(Y;X_{i})$. Similarly, \cref{eq:boundunion-1} implies
that the synergy, as defined in \cref{eq:synDefM}, obeys
\[
0\le S(X_{1};\dots;X_{n}\!\shortrightarrow\!Y)\le\min_{i}I(Y;X_{1},\dots,X_{n}\vert X_{i}),
\]
where we have used the chain rule $I(Y;X_{1},\dots,X_{n})=I(Y;X_{i})+I(Y;X_{1},\dots,X_{n}\vert X_{i})$.
\cref{eq:boundunion-1} also implies that excluded information in each
source $X_{i}$, as defined in\mbox{ \cref{eq:excdef}}, is bounded between
0 and $I(Y;X_{1},\dots,X_{n}\vert X_{i})$.

Note that in general, stronger orders give smaller values of redundancy
and larger values of union information. Consider two orders $\sqsubset$
and $\sqsubset'$ where the first one is stronger than the second:
$A\sqsubset B\implies A\sqsubset'B$ for all $A$ and $B$. Then,
any $Q$ in the feasible set of \cref{eq:IcapGen} under $\sqsubset$
will also be in the feasible set under $\sqsubset'$, and similarly
for \mbox{\cref{eq:IcupGen}}. Therefore, $I_{\cap}$ defined relative to
$\sqsubset$ will have a lower value than $I_{\cap}$ defined relative
to $\sqsubset'$, and vice versa for $I_{\cup}$.

In the rest of this section, we discuss alternative axiomatic justifications
for our general framework, the role of the inclusion-exclusion principle,
relation to prior work, and further generalizations. Readers who are
more interested in the use of our framework to define concrete measures
of redundancy and union information may skip to \cref{sec:partwo}.

\subsection{Axiomatic Derivation}

\label{subsec:Axiomatic-derivation}

In \cref{subsec:RedundancyDef}, we defined the PID in terms of an
algebraic analogy with intersection and union in set theory. This
definition can be considered as the primary one in our framework.
At the same time, the same definitions can also be derived in an alternative
manner from a set of axioms, as commonly sought after in the PID literature.
In particular, in \cref{app:unqproofs}, we prove the following result
regarding redundancy.
\begin{thm}
\label{thm:unq}Any redundancy measure that satisfies the following
five axioms is equal to  $I_{\cap}(X_{1};\dots;X_{n}\!\shortrightarrow\!Y)$
as defined in \cref{eq:IcapGen}.
\begin{enumerate}
\item \emph{\hspace{-10pt}Symmetry}: $I_{\cap}(X_{1};\dots;X_{n}\!\shortrightarrow\!Y)$
is invariant to the permutation of $X_{1},\dots,X_{n}$.
\item \emph{\hspace{-10pt}Self-redundancy}: $I_{\cap}(X_{1}\!\shortrightarrow\!Y)=I(Y;X_{1})$.
\item \emph{\hspace{-10pt}Monotonicity}: $I_{\cap}(X_{1};\dots;X_{n}\!\shortrightarrow\!Y)\le I_{\cap}(X_{1};\dots;X_{n-1}\!\shortrightarrow\!Y)$.
\item \emph{\hspace{-10pt}Order equality}: $I_{\cap}(X_{1};\dots;X_{n}\!\shortrightarrow\!Y)=I_{\cap}(X_{1};\dots;X_{n-1}\!\shortrightarrow\!Y)$
if $X_{i}\sqsubset X_{n}$ for some $i<n$.
\item \emph{\hspace{-10pt}Existence}:\emph{ }There is some $Q$ such that
$I_{\cap}(X_{1};\dots;X_{n}\!\shortrightarrow\!Y)=I(Y;Q)$ and $Q\sqsubset X_{i}$
for all $i$.
\end{enumerate}
\end{thm}

While \emph{Symmetry}, \emph{Self-redundancy}, and \emph{Monotonicity}
axioms are\textbf{ }standard in the PID literature (see \cref{app:axioms}),
the last two axioms require some explanation. \emph{Order equality}
is a generalization of the previously proposed \emph{Deterministic
equality} axiom, described in \cref{app:axioms}, where the condition
$X_{i}=f(X_{n})$ (deterministic relationship) is generalized to the
``more informative'' relation $X_{i}\sqsubset X_{n}$. This axiom
reflects the idea that if a new source $X_{n}$ is more informative
than an existing source $X_{i}$, then redundancy shouldn't decrease
when $X_{n}$ is added. 

\emph{Existence} is the most novel of our proposed axioms. It says
that for any set of sources $X_{1},\dots,X_{n}$, there exists some
random variable which captures the redundant information. It is similar
to the statement in axiomatic set theory that the intersection of
a collection of sets is itself a set (note that in Zermelo-Fraenkel set theory, this statement is derived from the Axiom of Separation).

We can derive a similar result for union information (proof in \cref{app:unqproofs}). 
\begin{thm}
\label{thm:unq-union}Any union information measure that satisfies
the following five axioms is equal to $I_{\cup}(X_{1};\dots;X_{n}\!\shortrightarrow\!Y)$
as defined in \cref{eq:IcupGen}.
\begin{enumerate}
\item \emph{\hspace{-10pt}Symmetry}: $I_{\cup}(X_{1};\dots;X_{n}\!\shortrightarrow\!Y)$
is invariant to the permutation of $X_{1},\dots,X_{n}$.
\item \emph{\hspace{-10pt}Self-union}: $I_{\cup}(X_{1}\!\shortrightarrow\!Y)=I(Y;X_{1})$.
\item \emph{\hspace{-10pt}Monotonicity}: $I_{\cup}(X_{1};\dots;X_{n}\!\shortrightarrow\!Y)\ge I_{\cup}(X_{1};\dots;X_{n-1}\!\shortrightarrow\!Y)$.
\item \emph{\hspace{-10pt}Order equality}: $I_{\cup}(X_{1};\dots;X_{n}\!\shortrightarrow\!Y)=I_{\cup}(X_{1};\dots;X_{n-1}\!\shortrightarrow\!Y)$
if $X_{n}\sqsubset X_{i}$ for some $i<n$.
\item \emph{\hspace{-10pt}Existence}:\emph{ }There is some $Q$ such that
$I_{\cup}(X_{1};\dots;X_{n}\!\shortrightarrow\!Y)=I(Y;Q)$ and $X_{i}\sqsubset Q$
for all $i$.
\end{enumerate}
\end{thm}

These axioms are dual to the redundancy axioms outlined above. Compared
to previously proposed axioms for union information, as described
in \cref{app:axioms}, the most unusual of our axioms is \emph{Existence}.
It says that given a set of sources $X_{1},\dots,X_{n}$, there exists
some random variable which captures the union information. It is similar
in spirit to the ``Axiom of Union'' in axiomatic set theory \citep{halmos2017naive}.

Finally, note that for some choices of $\sqsubset$, there may not
exist measures of redundancy and/or union information that satisfy
the axioms in \cref{thm:unq} and \cref{thm:unq-union}, in which case
these theorems still hold but are trivial. However, even in such ``pathological''
cases, $I_{\cap}$ and $I_{\cup}$ can still be defined via \cref{eq:IcapGen,eq:IcupGen}, 
as long as $\sqsubset$ has a ``least informative'' and a ``most
informative'' element (e.g., as provided by Assumption
III above), so that the feasible sets are not empty. In this sense,
the definitions in \cref{eq:IcapGen,eq:IcupGen} are more general
than the axiomatic derivations provided by \cref{thm:unq,thm:unq-union}.

\subsection{Inclusion-Exclusion Principle}

\label{sec:iep-violation}

One unusual aspect of our approach is that, unlike most previous work,
we propose separate measures of redundancy and union information.

Recall that in set theory, the size of the intersection and the union
are not independent of each other, but are instead related by the
\emph{inclusion-exclusion principle} (IEP). For example, given any
two sets $S$ and $T$, the IEP states that the size of the union
of $S$ and $T$ is given by the sum of their individual sizes minus
the intersection,
\begin{equation}
\left|S\cup T\right|=\left|S\right|+\left|T\right|-\left|S\cap T\right|.\label{eq:iepset}
\end{equation}
More generally, the IEP relates the sizes of intersection and unions
for any number of sets, via the following inclusion-exclusion formulas:
\begin{align}
\Big|\bigcup_{i=1}^{n}S_{i}\Big| & =\sum_{\varnothing\ne J\subseteq\{1,\dots,n\}}(-1)^{\left|J\right|-1}\Big|\bigcap_{i\in J}S_{i}\Big|.\label{eq:iepgen}\\
\Big|\bigcap_{i=1}^{n}S_{i}\Big| & =\sum_{\varnothing\ne J\subseteq\{1,\dots,n\}}(-1)^{\left|J\right|-1}\Big|\bigcup_{i\in J}S_{i}\Big|.\label{eq:iepunion}
\end{align}

Historically, the IEP has played an important role in analogies between
set theory and information theory, which began to be explored in 1950s
and 1960s \citep{mcgill1954multivariate,fanoTransmissionInformationStatistical1961,rezaIntroductionInformationTheory1961,ting1962amount,yeung1991new}.
Recall that the entropy $H(X)$ quantifies the amount of information
gained by learning the outcome of random variable $X$. It has been observed that,
for a set of random variables $X_{1},\dots,X_{n}$, the joint entropy
$H(X_{1},\dots,X_{n})$ behaves somewhat like the size of the union
of the information in the individual variables. For instance, like
the size of the union, joint entropy is subadditive ($H(X_{1})+H(X_{2})\ge H(X_{1},X_{2})$)
and increases with additional random variables ($H(X_{1},X_{2})\ge H(X_{1})$).
Moreover, for two random variables $X_{1}$ and $X_{2}$, the mutual
information $I(X_{1};X_{2})=H(X_{1})+H(X_{2})-H(X_{1},X_{2})$ acts
like the size of the intersection of the information provided by $X_{1}$
and $X_{2}$, once intersection is defined analogously to the IEP
expression in \cref{eq:iepset} \citep{ting1962amount,yeung1991new}.
Given the general IEP formula in \cref{eq:iepunion}, this can be used
to define the size of the intersection between any number of random
variables. For instance, the size of a three-way intersection is
\begin{align*}
I(X_{1};X_{2};X_{3}) & =H(X_{1})+H(X_{2})+H(X_{3})\\
 & \qquad-H(X_{1},X_{2})-H(X_{1},X_{3})-H(X_{2},X_{3})+H(X_{1},X_{2},X_{3}),
\end{align*}
a quantity called \emph{co-information }or \emph{interaction information}
in the literature \citep{fanoTransmissionInformationStatistical1961,mcgill1954multivariate,ting1962amount,yeung1991new,bell2003co}.

Unfortunately, interaction information, as well as other higher-order
interaction terms defined via the IEP, can take negative values \citep{ting1962amount,mcgill1954multivariate,bell2003co}.
This conflicts with the intuition that information measures should
always be non-negative, in the same way that set size is always non-negative.

One of the primary motivations for the PID, as originally proposed
by Williams and Beer \citep{williams2010nonnegative,williams_information_2011},
was to solve the problem of negativity encountered by interaction
information. To develop a non-negative information decomposition,
Williams and Beer took two steps. First, they considered the information
that a set of sources $X_{1},\dots,X_{n}$ provide about some target
random variable $Y$. Second, they developed a non-negative measure
of redundancy ($I_{\cap}^{\mathrm{WB}}$) which leads to a non-negative
union information once an IEP formula like \mbox{\cref{eq:iepgen}} is applied
\citep[Theorem 4.7, ][]{williams_information_2011}. For example,
in the original proposal, union information and redundancy are related
via
\begin{equation}
I_{\cup}(X_{1};X_{2}\!\shortrightarrow\!Y)\stackrel{?}{=}I(Y;X_{1})+I(Y;X_{2})-I_{\cap}(X_{1};X_{2}\!\shortrightarrow\!Y),\label{eq:iep-1}
\end{equation}
\textls[-10]{which is the analogue of \cref{eq:iepset}. This can be plugged into
expressions like \mbox{\cref{eq:synDefM}},} so as to express synergy in terms
of redundancy as
\begin{equation}
S(X_{1};\dots;X_{n}\!\shortrightarrow\!Y)\stackrel{?}{=}I(Y;X_{1},\dots,X_{n})-I(Y;X_{1})-I(Y;X_{2})+I_{\cap}(X_{1};X_{2}\!\shortrightarrow\!Y).\label{eq:syniep}
\end{equation}
The meaning of IEP-based identities such as \cref{eq:iep-1,eq:syniep}
can be illustrated using the Venn diagrams in \cref{fig:venn}. In
particular, they imply that the pink region in the right diagram is
equal in size to the pink region in the left diagram, and that the
grey region in the left diagram is equal in size to the grey region
in the right diagram. More generally, IEP implies an equivalence between
the information decomposition based on redundancy and the one based
on union information.

As mentioned in \cref{sec:Background}, due to shortcomings in the
original redundancy measure $I_{\cap}^{\mathrm{WB}}$, numerous other
proposals for the PID have been advanced. Most of these proposals
introduce new measures of redundancy, while keeping the general structure
of the PID as introduced by Williams and Beer. In particular, most
of these proposals assume that the IEP holds, so that union information
can be derived from a measure of redundancy. While the assumption
of the IEP is sometimes stated explicitly, more frequently it is implicit
in the definitions used. For example, many proposals assume that synergy
is related to redundancy via an expression like \cref{eq:syniep},
although (as shown above) this implicitly assumes that the IEP holds.
In general, the IEP has been largely an unchallenged and unexamined
assumption in the PID field. It is easy to see the appeal of the IEP:
it builds on deep-seated intuitions about intersection/union from set theory and
Venn diagrams, it has a long history in the information-theoretic
literature, and it simplifies the problem of defining the PID since
it only requires a measure of redundancy to be defined --- rather than
a measure of redundancy and a measure of union information. (Note that one can also start from union information and then derive
 redundancy via the IEP formula in \cref{eq:iepunion}, as in Appendix~B of Ref.~\cite{griffith2014quantifying},
although this is much less common in the literature.)

However, there is a different way to define a non-negative PID, which
is still grounded in a formal analogy with set theory but does not
assume the IEP. Here, one defines measures of redundancy and union
information based on the underlying algebra of intersection and union:
the intersection of $X_{1},\dots,X_{n}$ is the largest element that
is less than each $X_{i}$, while the union is the smallest element
that is greater than each $X_{i}$. Given these definitions, intersections
and unions are not necessarily related to each numerically, as in
the IEP, but are instead related by an algebraic duality. 

This latter approach is the one we pursue in our definitions (it has
also appeared in some prior work, which we review in the next subsection).
In general, the IEP will not hold for redundancy and union information
as defined in \cref{eq:IcapGen,eq:IcupGen}. (To emphasize
this point, we put a question mark in \cref{eq:iep-1,eq:syniep},
and made the sizes of the pink and grey regions visibly different
in \cref{fig:venn}.) However, given the algebraic and axiomatic justifications
for $I_{\cap}$ and $I_{\cup}$, we do not see the violation of the
IEP as a fatal issue. In fact, there are many domains where generalizations
of intersections and unions do not obey the IEP. For example, it is
well-known that the IEP is violated in the domain of vector spaces,
once the size of a vector space is measured in terms of its dimension
\citep{iepthreevectorMO}. The PID is simply another domain where
the IEP should not be expected to hold.

We believe that many problems encountered in previous work on the
PID --- such as the failure of certain redundancy measures to generalize
to more than two sources, or the appearance of uninterpretable negative
synergy values --- are artifacts of the IEP assumption. In fact,
the following result shows that any measures of redundancy and union
information which satisfy several reasonable assumptions must violate
the IEP as soon as 3 or more sources are present (the proof, in \cref{app:misc},
is based on a construction from \citep{rauh2014reconsidering,rauh_secret_2017}).
\begin{lem}
\label{lem:iep}\textls[-15]{Let $I_{\cap}$ be any nonnegative redundancy measure
which obeys \emph{Symmetry}, \emph{Self-redundancy}}, \emph{Monotonicity},
and \emph{Independent identity}. Let $I_{\cup}$ be any union information
measure which obeys $I_{\cup}(X_{1};\dots;X_{n}\!\shortrightarrow\!Y)\le I(Y;X_{1},\dots,X_{n})$.
Then, $I_{\cap}$ and $I_{\cup}$ cannot be related by the inclusion-exclusion
principle for 3 or more sources.
\end{lem}

The idea that different information decompositions
may arise from redundancy versus synergy (and therefore union information)
has recently appeared in the PID literature \citep{james2018unique,rauh_secret_2017,chicharro_redundancy_2016,ay2019information,rosas2020operational}.
In particular, Chicharro and Panzeri proposed a PID that involves
two decomposition: an ``information gain'' decomposition based on
redundancy and an ``information loss'' decomposition based on synergy
\citep{chicharro_redundancy_2016}. These decompositions correspond
to the two Venn diagrams shown in \cref{fig:venn}.

\subsection{Relation to Prior Work}

\label{subsec:Relation-to-prior}

Here we discuss prior work which is relevant to our algebraic approach
to the PID.

First, note that our definitions of redundancy and union information
in  {\cref{eq:IcapGen,eq:IcupGen}} are closely related to notions of
``meet'' and ``join'' in a field of algebra called order theory,
which generalize intersections and unions to domains beyond set theory
\citep{daveyIntroductionLatticesOrder2002}. Given a set of objects
$S$ and an order $\sqsubset$, the \emph{meet} of $a,b\in S$ is
the unique largest $c\in S$ that is smaller than both $a$ and $b$:
$c\sqsubset a,c\sqsubset b$ and $d\sqsubset c$ for any $d$ that
obeys $d\sqsubset a,d\sqsubset b$. Similarly, the \emph{join} of
$a,b\in S$ is the unique smallest $c$ that is larger than both $a$
and $b$: $a\sqsubset c,a\sqsubset c$ and $c\sqsubset d$ for any
$d$ that obeys $a\sqsubset d,b\sqsubset d$. Note that meets and
joins are only defined when $\sqsubset$ is a special type of partial
order called a \emph{lattice. }This is a strict requirement, and many
important ordering relations in information theory are not lattices
(this includes the ``Blackwell order'', which we will consider in
part II of this paper \citep{bertschinger2014blackwell}).

In our approach, we do not require the ordering relation $\sqsubset$
to be a lattice, or even a partial order. We do not require these
properties because we do not aim to find the unique union random variable
or the unique redundancy random variable. Instead, we aim to quantify the
\emph{size of the intersection} and \emph{the size of the union},
which we do by optimizing mutual information subject to constraints,
as \cref{eq:IcapGen,eq:IcupGen}. These definitions are well-defined
even when $\sqsubset$ is not a lattice, which allows us to consider
a much broader set of ordering relations.

We mention three important precursors of our approach that have been
proposed in the PID literature. First, Griffith et al. \citep{griffith2014intersection}
considered the following order between random variables:
\begin{equation}
A\vartriangleleft B\text{ iff \ensuremath{A=f(B)} for some deterministic function \ensuremath{f}.}\label{eq:funcrel-1}
\end{equation}
This ordering relation $\vartriangleleft$ was first considered in
a 1953 paper by Shannon \citep{shannon_lattice_1953}, who showed
that it defines a lattice over random variables. That paper was the
first to introduce the algebraic idea of meets and joins into information
theory, leading to an important line of subsequent research~\citep{li2011connection,gacs_common_1973,aumann1976agreeing,banerjee2015synergy,hexner1977information}.
Using this order, Ref.~\citep{griffith2014intersection} defined redundancy
as the maximum mutual information in any random variable that is a
deterministic function of all of the sources,
\begin{equation}
I_{\cap}^{\vartriangleleft}(X_{1};\dots;X_{n}\!\shortrightarrow\!Y):=\max_{Q}I(Q;Y)\quad\text{such that}\quad\forall i\;Q\vartriangleleft X_{i},\label{eq:IwedgeDef}
\end{equation}
which is clearly a special case of \cref{eq:IcapGen}. Unfortunately,
in practice, $I_{\cap}^{\vartriangleleft}$ is not a useful redundancy
measure, as it tends to give very small values and is highly discontinuous.
For example, $I_{\cap}^{\vartriangleleft}(X_{1};\dots;X_{n}\!\shortrightarrow\!Y)=0$
whenever the joint distribution $P_{X_{1}\dots X_{n}Y}$ has full
support, meaning that it vanishes on almost all joint distributions
\citep{gacs_common_1973,griffith2014intersection,griffith_quantifying_2015}.
The reason for this counterintuitive behavior is that the order $\vartriangleleft$
formalizes an extremely strict notion of ``more informative'', which
is not robust to noise.

Given the deficiencies of $I_{\cap}^{\vartriangleleft}$, Griffith
and Ho \citep{griffith_quantifying_2015} proposed another measure
of redundancy (also discussed as $I_{\cap}^{2}$ in Ref.~\citep{banerjee2015synergy}),
\begin{equation}
I_{\cap}^{\mathrm{GH}}(X_{1};\dots;X_{n}\!\shortrightarrow\!Y):=\max_{Q}I(Q;Y)\quad\text{such that}\quad\forall i\;Q-X_{i}-Y.\label{eq:igh}
\end{equation}
This measure is also a special case of \cref{eq:IcapGen}, where the
more informative relation $A\sqsubset B$ is formalized via the conditional
independence condition $A-B-Y$. This measure is similar to the redundancy
measure we propose in part II of this paper, and we discuss it in
more detail in \cref{subsec:Blackwell-union-information-priorwork}. 
(Note that there are some incorrect claims about $I_{\cap}^{\mathrm{GH}}$
in the literature: Lemmas 6 and 7 of Ref.~\citep{banerjee2015synergy}
incorrectly state that $I_{\cap}^{\mathrm{GH}}(X_{1};X_{2}\!\shortrightarrow\!Y)=0$
whenever $X_{1}$ and $X_{2}$ are independent --- see the AND gate
counterexample in \cref{sec:Relation-to-existing} --- while Ref.~\citep{griffith_quantifying_2015}
incorrectly states that $I_{\cap}^{\mathrm{GH}}$ obeys a property
called \emph{Target Monotonicity}.)

Finally, we mention the so-called ``minimum mutual information''
redundancy $I_{\cap}^{\text{\ensuremath{\mathrm{MMI}}}}$ \citep{barrett_exploration_2015}.
This is perhaps the simplest redundancy measure, being equal to the
minimal mutual information in any source: $I_{\cap}^{\text{\ensuremath{\mathrm{MMI}}}}(X_{1};\dots;X_{n}\!\shortrightarrow\!Y):=\min_{i}I(X_{i};Y)$.
It can be written in the form of \cref{eq:IcapGen} as
\begin{equation}
I_{\cap}^{\text{\ensuremath{\mathrm{MMI}}}}(X_{1};\dots;X_{n}\!\shortrightarrow\!Y):=\max_{Q}I(Q;Y)\quad\text{such that}\quad\forall i\;I(Q;Y)\le I(X_{i};Y).\label{eq:mmi}
\end{equation}
This redundancy measure has been criticized for depending only on
the amount of information provided by the different sources, being
completely insensitive to the content of that information. Nonetheless,
$I_{\cap}^{\text{\ensuremath{\mathrm{MMI}}}}$ can be useful in some
settings, and it plays an important role in the context of Gaussian
random variables \citep{barrett_exploration_2015}.

Interestingly, unlike $I_{\cap}^{\text{\ensuremath{\mathrm{MMI}}}}$,
the original redundancy measure proposed by Williams and Beer \citep{williams2010nonnegative},
$I_{\cap}^{\mathrm{WB}}$, does not appear to be a special case of
\cref{eq:IcapGen} (at least not under the natural definition of the
ordering relation $\sqsubset$). We demonstrate this using a counter-example
in \cref{app:IwbGen}.

As mentioned in \cref{subsec:RedundancyDef}, stronger ordering relations
give smaller values of redundancy. For the orders considered above,
it is easy to show that
\begin{equation}
A\vartriangleleft B\implies A-B-Y\implies I(A;Y)\le I(B;Y).\label{eq:order2}
\end{equation}
This implies that $I_{\cap}^{\vartriangleleft}\le I_{\cap}^{\mathrm{GH}}\le I_{\cap}^{\text{\ensuremath{\mathrm{MMI}}}}$
. In fact, $I_{\cap}^{\text{\ensuremath{\mathrm{MMI}}}}$ is the largest
measure that is compatible with the monotonicity of mutual information
(Assumption I in \mbox{\cref{subsec:RedundancyDef}).}

\subsection{Further Generalizations}

\label{subsec:Further-generalizations}

We finish part I of this paper by noting that one can further generalize
our approach, by considering other analogues of ``set'', ``set size'',
and ``set inclusion'' beyond the ones considered in \cref{subsec:RedundancyDef}.
Such generalizations allow one to analyze notions of information intersection
and union in a wide variety of domains, including setups different
from the standard one considered in the PID, and domains not based
on Shannon information theory.

At a general level, consider a set of object $\Omega$ that represents
possible ``sources'', which may be random variables, as in \cref{subsec:RedundancyDef},
or otherwise. Assume there is some function $\phi:\Omega\to\mathbb{R}$
that quantifies the ``amount of information'' in a given source
$\Omega$, and some relation $\sqsubset$ on $\Omega$ that indicates
which sources are more informative than others. Then, in analogy to
\cref{eq:int0,eq:union0-1}, for any set of sources $\{b_{1},\dots,b_{n}\}\subseteq\Omega$,
one can define redundancy and union information as 
\begin{align}
I_{\cap}(b_{1};\dots;b_{n}) & :=\sup_{a\in\Omega}\;\phi(a)\quad\text{such that}\quad\forall i\;a\sqsubset b_{i}\label{eq:genMeet}\\
I_{\cup}(b_{1};\dots;b_{n}) & :=\inf_{a\in\Omega}\;\phi(a)\quad\text{such that}\quad\forall i\;b_{i}\sqsubset a.\label{eq:genJoin}
\end{align}
Synergy, unique, and excluded information can then be defined via
\cref{eq:unqdef,eq:excdef,eq:synDefM}.

There are many possible examples of such generalizations, of which
we mention a few as illustrations.

\vspace{5pt}
\noindent \emph{Shannon information theory (beyond mutual information)}. In
\cref{subsec:RedundancyDef}, $\phi$ was the mutual information between
each random variable and some target $Y$. This can be generalized
by choosing a different ``amount of information'' function $\phi$,
so that redundancy and union information are quantified in terms of
other measures of statistical dependence. Among many other options,
possible choices of $\phi$ include Pearson's correlation (for continuous
random variables) and measures of statistical dependency based $f$-divergences
\citep{pluim2004f}, Bregman divergences \citep{banerjee2005clustering},
and Fisher information \citep{brunel1998mutual}.

\vspace{5pt}
\noindent \emph{Shannon information theory (without a fixed target)}. The PID
can also be defined for a different setup than the typical one considered
in the literature. For example, consider a situation where the sources
are channels $\kappa_{X_{1}|Y},\dots,\kappa_{X_{n}\vert Y}$, while
the marginal distribution over the target $Y$ is left unspecified.
Here one may take $\Omega$ as the set of channels, $\phi$ as the
channel capacity $\phi(\kappa_{A\vert Y}):=\max_{P_{Y}}I_{P_{Y}\kappa_{A|Y}}(A;Y)$,
and $\sqsubset$ as some ordering relation on channels \citep{cohen1998comparisons}

\vspace{5pt}
\noindent \emph{Algorithmic information theory.} The PID can be defined for
other notions of information, such as the ones used in Algorithmic
Information Theory (AIT) \citep{li2008introduction}. In AIT, ``information''
is not defined in terms of statistical uncertainty, but rather in
terms of the program length necessary to generate strings. For example,
one may take $\Omega$ as the set of finite strings, $\sqsubset$
as algorithmic conditional independence ($a\sqsubset b\text{\,iff\,}K(y\vert b)-K(y\vert b,a)\le\text{const}$,
where $K(\cdot\vert\cdot)$ is conditional Kolmogorov complexity),
and $\phi(a):=K(y)-K(y\vert a)$ as the ``algorithmic mutual information''
with some target string $y$. (This setup is closely related
to the notion of algorithmic ``common information'' \citep{gacs_common_1973}.)

\vspace{5pt}
\noindent \emph{Quantum information theory}. As a final example, the PID can
be defined in the context of quantum information theory. For example,
one may take $\Omega$ as the set of quantum channels, $\sqsubset$
as quantum Blackwell order \citep{shmaya2005comparison,chefles2009quantum,buscemi2012comparison},
and $\phi(\Phi)=\mathcal{I}(\rho,\Phi)$, where $\mathcal{I}$ is
the Ohya mutual information for some target density matrix $\rho$
under channel $\Phi\in \Omega$ \citep{ohya2010quantum}.

\section{Part II: Blackwell Redundancy and Union Information}

\label{sec:partwo}

In the first part of this paper, we proposed a general framework for
defining PID terms. In this section, which forms part II of this paper,
we develop a concrete definition of redundancy and union information
by combining our general framework with a particular ordering relation
$\sqsubset$. This ordering relation is called the ``Blackwell order'',
and it plays a fundamental role in statistics and decision theory
\citep{blackwell_equivalent_1953,bertschinger2014blackwell,rauh_coarse-graining_2017}.
We first introduce the Blackwell order, then use it to define measures
of redundancy and union information, and finally discuss various properties
of our measures.

\subsection{The Blackwell Order}

\label{subsec:The-Blackwell-order}

We begin by introducing the ordering relation that we use to define
our PID. Given three random variables $B,C$ and $Y$, the ordering
relation $B\prec_{Y}C$ is defined as follows:
\begin{align}
B\prec_{Y}C\;\;\text{iff}\;\;P_{B|Y}(b\vert y) & =\sum_{c}\kappa_{B\vert C}(b\vert c)P_{C|Y}(c\vert y)\;\;\text{for some channel \ensuremath{\kappa_{B\vert C}} and all \ensuremath{b,y}}.\label{eq:garbl}
\end{align}
We refer to the relation $\prec_{Y}$ as the \emph{Blackwell order}
relative to random variable $Y$. (Note that the Blackwell order and
Blackwell's Theorem are usually formulated in terms of channels ---
that is, conditional distributions like $\kappa_{B\vert Y}$ and $\kappa_{C\vert Y}$
--- rather than of random variables as done here. However, these
two formulations are equivalent, as shown in \citep{bertschinger2014blackwell}.)

In words, \cref{eq:garbl} means the conditional distribution by $P_{B|Y}$
can be generated by first sampling from the conditional distribution
$P_{C|Y}$, and then applying some channel $\kappa_{B\vert C}$ to
the outcome. The relation $B\prec_{Y}C$ implies that $P_{B|Y}$ is
more noisy than $P_{C|Y}$ and, by the ``data processing inequality''
\citep{cover_elements_2006},  $B$ must have less mutual
information about $Y$ than $C$:
\begin{align}
B & \prec_{Y}C\implies I(B;Y)\le I(C;Y).\label{eq:dpi}
\end{align}

Intuition suggests that when $B\prec_{Y}C$, the information that
$B$ provides about $Y$ is contained in the information that $C$
provides about $Y$. This intuition is formalized within a decision-theoretic
framework using the so-called Blackwell's Theorem \citep{blackwell_equivalent_1953,bertschinger2014blackwell,rauh_coarse-graining_2017}.
To introduce this theorem, imagine a scenario in which $Y$ represents
the state of the environment. Imagine also that there is an agent
who acquires information about the environment via the conditional
distribution $P_{B|Y}(b\vert y)$, and then uses outcome $B=b$ to
select actions $a\in\mathcal{A}$ according to some ``decision rule''
given by the channel $\kappa_{A\vert B}$. Finally, the agent gains
utility according to some utility function $u(a,y)$, which depends
on the agent's action $a$ and the environment's state $y$. The maximum
expected utility achievable by any decision rule is given by 
\begin{equation}
V_{Y}^{\max}(B,u):=\max_{\kappa_{A\vert B}}\sum_{y,b,a}P_{Y}(y)P_{B\vert Y}(b\vert y)\kappa_{A\vert B}(a\vert b)u(a,y).\label{eq:eu}
\end{equation}
From an operational perspective, it is natural to say that $B$ is
less informative than $C$ about $Y$ if there is no utility function
such that an agent with access to $B$ can achieve higher expected
utility than an agent with access to $C$. Blackwell's Theorem states
that this is precisely the case if and only if $B\prec_{Y}C$ \citep{blackwell_equivalent_1953,bertschinger2014blackwell}:
\begin{equation}
B\prec_{Y}C\quad\text{iff}\quad V_{Y}^{\max}(B,u)\le V_{Y}^{\max}(C,u)\quad\text{for all \ensuremath{u}.}\label{eq:blthm}
\end{equation}
In some sense, this operational description of the relation $\prec_{Y}$
is deeper than the data processing inequality, \cref{eq:dpi}, which says that $B\prec_{Y}C$
is sufficient (but not necessary) for $I(B;Y)\le I(C;Y)$. In
fact, it can happen that $I(B;Y)\le I(C;Y)$ even though $B\not\prec_{Y}C$
\citep{korner1977comparison,rauh_coarse-graining_2017,makur2018comparison}.

A connection between PID and Blackwell's theorem was first proposed
in \citep{bertschinger2014quantifying}, which argued that the PID
should be defined in an operational manner (see \cref{subsec:Blackwell-union-information}
for further discussion of \citep{bertschinger2014quantifying}).

\subsection{Blackwell Redundancy}

We now define a measure of redundancy based on the Blackwell order.
Specifically, we use our general definition of redundancy, \cref{eq:IcapGen},
while using the Blackwell order relative to $Y$ as the ``more informative''
relation $\sqsubset$:
\begin{equation}
I_{\cap}^{\prec}(X_{1};\dots;X_{n}\!\shortrightarrow\!Y):=\sup_{Q}\;I(Q;Y)\quad\text{such that}\quad\forall i\;Q\prec_{Y}X_{i}.\label{eq:istarmaintext}
\end{equation}
We refer to this measure as \emph{Blackwell redundancy}.

Given Blackwell's Theorem, $I_{\cap}^{\prec}$ has a simple operational
interpretation. Imagine two agents, Alice and Bob, who can acquire
information about $Y$ via different random variables, and then use
this information to maximize their expected utility. Suppose that
Alice has access to one of the sources $X_{i}$. Then, the Blackwell
redundancy $I_{\cap}^{\prec}$ is the maximum information that Bob
can have about $Y$ without being able to do better than Alice on
any utility function, regardless of which source Alice has access
to.

Blackwell redundancy can also be used to define a measure of Blackwell
unique information, $U^{\prec}(X_{i}\!\shortrightarrow\!Y\vert X_{1};\dots;X_{n}):=I(Y;X_{i})-I_{\cap}^{\prec}(X_{1};\dots;X_{n}\!\shortrightarrow\!Y)$,
via \cref{eq:unqdef}. As we show in \cref{app:misc}, $U^{\prec}$
satisfies the following property, which we term the \emph{Multivariate
Blackwell property}.
\begin{thm}
\label{thm:blackwell}$U^{\prec}(X_{i}\!\shortrightarrow\!Y\vert X_{1};\dots;X_{n})=0$
if and only if $X_{i}\prec_{Y}X_{j}$ for all $j\ne i$.
\end{thm}

Operationally, \cref{thm:blackwell} means that source $X_{i}$ has
non-zero unique information iff there exists a utility function such
that an agent with access to source $X_{i}$ can achieve higher utility
than an agent with access to any other source $X_{j}$.

Computing $I_{\cap}^{\prec}$ involves maximizing a convex function
subject to a set of linear constraints. These constraints define a
feasible set which is a convex polytope, and the maximum must lie
on one of the vertices of this polytope \citep{benson1995concave}.
 In \cref{app:opt}, we show how to solve this optimization problem.
In particular, we use a computational geometry package to enumerate
the vertices of the feasible set, and then choose the best vertex
(code is available at \citep{githubRepo}). In that appendix, we also
prove that an optimal solution to \mbox{\cref{eq:istarmaintext}} can always
be achieved by $Q$ with cardinality $\left|\mathrm{\mathcal{Q}}\right|=(\sum_{i}\left|\mathcal{\mathcal{X}}_{i}\right|)-n+1$.
Note that the supremum in \cref{eq:istarmaintext} is
always achieved. Note also that $I_{\cap}^{\prec}$
satisfies the redundancy axioms in \cref{subsec:Axiomatic-derivation}.

As discussed above, solving the optimization problem in \cref{eq:istarmaintext} gives a (possibly non-unique) optimal random variable $Q$ which
specifies the content of the redundant information. As shown in \cref{app:opt},
solving \cref{eq:istarmaintext} also provides a set of channels $\kappa_{Q\vert X_{i}}$
for each source $X_{i}$, which identify the redundant information
in each source.

Note that the Blackwell order satisfies assumptions I-III in \cref{subsec:RedundancyDef},
thus Blackwell redundancy satisfies the bounds derived in that section.
Finally, note that like many other redundancy measures, Blackwell
redundancy becomes equivalent to the measure $I_{\cap}^{\text{\ensuremath{\mathrm{MMI}}}}$
(as defined in \cref{eq:mmi}) when applied to Gaussian random variables
(for details, see \cref{app:gaussian}).

\subsection{Blackwell Union Information}

\label{subsec:Blackwell-union-information}

We now define a measure of union information using our general definition
in \cref{eq:IcupGen}, while using the Blackwell order relative
to $Y$ as the ``more informative'' relation:
\begin{equation}
I_{\cup}^{\prec}(X_{1};\dots;X_{n}\!\shortrightarrow\!Y):=\inf_{Q}\;I(Q;Y)\quad\text{such that}\quad\forall i\;X_{i}\prec_{Y}Q.\label{eq:ustar}
\end{equation}
We refer to this measure as \emph{Blackwell union information}.

As for Blackwell redundancy, Blackwell union information can be understood
in operational terms. Consider two agents, Alice and Bob, whose use
information about $Y$ to maximize their expected utility. Suppose
that Alice has access to one of the sources $X_{i}$. Then, the Blackwell
union information $I_{\cup}^{\prec}$ is the minimum information that
Bob must have about $Y$ in order to do better than Alice on any utility
function, regardless of which source Alice has access to.

Blackwell union information can be used to define measures of synergy
and excluded information via \cref{eq:excdef,eq:synDefM}. The resulting
measure of excluded information $E^{\prec}(X_{i}\!\shortrightarrow\!Y\vert X_{1};\dots;X_{n}):=I_{\cup}^{\prec}(X_{1};\dots;X_{n}\!\shortrightarrow\!Y)-I(Y;X_{i})$
satisfies the following property, which is the ``dual''
of the \emph{Multivariate Blackwell property} considered in \cref{thm:blackwell}.
(See \cref{app:misc} for the proof.)
\begin{thm}
\label{thm:blackwell-union}$E^{\prec}(X_{i}\!\shortrightarrow\!Y\vert X_{1};\dots;X_{n})=0$
if and only if $X_{j}\prec_{Y}X_{i}$ for all $j\ne i$.
\end{thm}

Operationally, \cref{thm:blackwell-union} means that there is excluded information
for source $X_{i}$ iff there exists a utility function such
that an agent with access to one of the other sources $X_{j}$ can
achieve higher expected utility than an agent with access to $X_{i}$.

We discuss the problem of numerically solving the optimization problem
in  \cref{eq:ustar} in the next subsection.

\subsection{Relation to Prior Work}

\label{subsec:Blackwell-union-information-priorwork}

Our measure of Blackwell redundancy $I_{\cap}^{\prec}$ is new to
the PID literature. The most similar existing redundancy measure is
$I_{\cap}^{\mathrm{GH}}$ \citep{griffith_quantifying_2015}, which
is discussed above in \cref{subsec:Relation-to-prior}. $I_{\cap}^{\mathrm{GH}}$
is a special case of \cref{eq:IcapGen}, once the ``more informative''
relation $B\sqsubset C$ is defined in terms of conditional independence
$B-C-Y$. Note that conditional independence is stronger than the
Blackwell order: given the definition of $\prec_{Y}$ in \cref{eq:garbl},
it is clear that $B-C-Y$ implies $B\prec_{Y}C$ (the channel $\kappa_{B\vert C}$
can be taken to be $P_{B\vert C}$), but not vice versa. As discussed
in \cref{subsec:RedundancyDef}, stronger ordering relations give smaller
values of redundancy, so in general $I_{\cap}^{\mathrm{GH}}\le I_{\cap}^{\prec}$
. Note also that $B\prec_{Y}C$ depends only on the pairwise marginals
$P_{BY}$ and $P_{CY}$, while conditional independence $B-C-Y$ depends
on the joint distribution $P_{BCY}$. As we discuss in \cref{app:Operational-interpretation-of-GH},
the conditional independence order can be interpreted in decision-theoretic terms, which suggests an operational interpretation
for $I_{\cap}^{\mathrm{GH}}$.

Interestingly, Blackwell union information $I_{\cup}^{\prec}$ is
equivalent to two measures that have been previously proposed in the
PID literature, although they were formulated in a different way.
Bertschinger et al. \citep{bertschinger2014quantifying} considered
the following measure of bivariate redundancy:
\begin{equation}
I_{\cap}^{\mathrm{BROJA}}(X_{1};X_{2}\!\shortrightarrow\!Y):=I(Y;X_{1})+I(Y;X_{2})-I_{\cup}^{\mathrm{BROJA}}(X_{1};X_{2}\!\shortrightarrow\!Y),\label{eq:broja1}
\end{equation}
where $I_{\cup}^{\mathrm{BROJA}}$ is defined via the optimization
problem
\begin{equation}
I_{\cup}^{\mathrm{BROJA}}(X_{1};X_{2}\!\shortrightarrow\!Y)=\min_{\tilde{X}_{1},\tilde{X}_{2}}I(Y;\tilde{X}_{1},\tilde{X}_{2})\;\text{such that}\; P_{\tilde{X}_{1}Y}=P_{X_{1}Y},P_{\tilde{X}_{2}Y}=P_{X_{2}Y},\label{eq:brojaopt}
\end{equation}
and reflects the minimal mutual information that two random variables
can have about $Y$, given that their pairwise marginals with $Y$
are fixed to be $P_{X_{1}Y}$ and $P_{X_{2}Y}$. Note that Ref.~\citep{bertschinger2014quantifying}
did not refer to $I_{\cup}^{\mathrm{BROJA}}$ as a measure of union
information (we use our notation in writing it as $I_{\cup}^{\mathrm{BROJA}}$).
Instead, these measures were derived from an operational motivation,
with the goal of deriving a unique information measure that obeys
the so-called \emph{Blackwell property}: $I(Y;X_{1})-I_{\cap}^{\mathrm{BROJA}}(X_{1};X_{2}\!\shortrightarrow\!Y)=0$
if $X_{1}\prec_{Y}X_{2}$ (see \cref{thm:blackwell,thm:blackwell-union}
above).

Starting from a different motivation, Griffith and Koch \citep{griffith2014quantifying}
proposed a multivariate version of $I_{\cup}^{\mathrm{BROJA}}$,
\begin{equation}
I_{\cup}^{\mathrm{BROJA}}(X_{1};\dots;X_{n}\!\shortrightarrow\!Y)=\min_{\tilde{X}_{i},\dots,\tilde{X}_{n}}I(Y;\tilde{X}_{1},\dots,\tilde{X}_{n})\;\text{such that}\;\forall i\;P_{\tilde{X}_{i}Y}=P_{X_{i}Y}.\label{eq:brojaGenOpt}
\end{equation}
The goal of Ref.~\citep{griffith2014quantifying} was to derive a
measure of multivariate synergy from a measure of union information,
as in \cref{eq:synDefM}. In that paper, $I_{\cup}^{\mathrm{BROJA}}$
was explicitly defined as a measure of union information. To our knowledge,
Ref.~\citep{griffith2014quantifying} was the first (and perhaps
only) paper to propose a measure of union information that was not
derived from redundancy via the inclusion-exclusion principle.

While $I_{\cup}^{\mathrm{BROJA}}(X_{1};\dots;X_{n}\!\shortrightarrow\!Y)$ and $I_{\cup}^{\prec}(X_{1};\dots;X_{n}\!\shortrightarrow\!Y)$
are stated as different optimization problems, we prove in \cref{app:brojaequivalent}
that these optimization problems are equivalent, in that they will
always achieve the same optimum value. Interestingly, since $I_{\cup}^{\mathrm{BROJA}}$
and $I_{\cup}^{\prec}$ are equivalent, our measure of Blackwell redundancy
$I_{\cap}^{\prec}$ appears as the natural dual to $I_{\cup}^{\mathrm{BROJA}}$.
Another implication of this equivalence is that Blackwell union information
$I_{\cup}^{\prec}$ can be quantified by solving the optimization
problem in \cref{eq:brojaGenOpt}, rather than \mbox{\cref{eq:ustar}.} This
is advantageous, because \cref{eq:brojaGenOpt} involves the minimization
of a convex function over a convex polytope, which can be solved using
standard convex optimization techniques \citep{banerjee2018computing}.

In Ref.~\citep{bertschinger2014quantifying}, the redundancy measure
$I_{\cap}^{\mathrm{BROJA}}$ in \cref{eq:broja1} was only defined
for the bivariate case. Since then, it has been unclear how to extend
this redundancy measure to more than two sources. However, by comparing
\cref{eq:broja1} and \cref{eq:iep-1}, we see the root of the problem:
$I_{\cap}^{\mathrm{BROJA}}$ is derived by applying the inclusion-exclusion
principle to a measure of union information, $I_{\cup}^{\mathrm{BROJA}}$.
It cannot be extended to more than two sources because the inclusion-exclusion
principle generally leads to counterintuitive results for more than
2 sources, as shown in \cref{lem:iep}. Note also that what Ref.~\citep{bertschinger2014quantifying} called the unique information
in $X_{1}$, $I_{\cup}^{\mathrm{BROJA}}(X_{1};X_{2}\!\shortrightarrow\!Y)-I(Y;X_{2})$,
in our framework would be considered a measure of the excluded information
for $X_{2}$.

At the same time, the union information measure $I_{\cup}^{\mathrm{BROJA}}$,
and the corresponding synergy from \cref{eq:synDefM}, does not use
the inclusion-exclusion principle. Therefore, it can be easily extended
to any number of sources \citep{griffith2014quantifying}.

\subsection{Continuity of Blackwell Redundancy and Union Information}

It is often desired that information-theoretic measures are continuous,
meaning that small changes in underlying probability distributions
lead to small changes in the resulting measures. In this section,
we consider the continuity of our proposed measures, $I_{\cap}^{\prec}$
and $I_{\cup}^{\prec}$.

We first consider Blackwell redundancy $I_{\cap}^{\prec}$. It turns
out that this measure is not always continuous in the joint probability
$P_{X_{1}\dots X_{n}Y}$ (a discontinuous example is provided in \cref{sec:identity}).
However, the discontinuity of $I_{\cap}^{\prec}$ is not necessarily
pathological, and we can derive an interpretable geometric condition
that guarantees that $I_{\cap}^{\prec}$ is continuous. 

Consider the conditional distribution of the target $Y$ given some
source $X_{i}$, $P_{Y\vert X_{i}}$. Let $\mathrm{rank\;}P_{Y\vert X_{i}}$
indicate its \emph{rank}, meaning the dimension of the space spanned
by the vectors $\{P_{Y\vert X_{i}=x_{i}}\}_{x_{i}\in\mathcal{X}_{i}}$.
The rank of $P_{Y\vert X_{i}}$ quantifies the number of independent
directions that the target distribution $P_{Y}$ can be moved by manipulating
the source distribution $P_{X_{i}}$, and it cannot be larger than $|\mathrm{\mathcal{Y}}|$. The next
theorem shows that $I_{\cap}^{\prec}$ is locally continuous, as long
as $n-1$ or more of the source conditional distributions have this maximal  
 rank.
\begin{thm}
\label{thm:cont}As a function of the joint distribution $P_{X_{1},\dots,X_{n},Y}$,
$I_{\cap}^{\prec}$ is locally continuous whenever $n-1$ or more
of the conditional distributions $P_{Y\vert X_{i}}$ have $\mathrm{rank\;}P_{Y\vert X_{i}}=|\mathrm{\mathcal{Y}}|$.
\end{thm}

In proving this result, we also show that $I_{\cap}^{\prec}$ is continuous
almost everywhere (see proof in \cref{app:continuity}). Finally, in
that appendix we also use \cref{thm:cont} to show that $I_{\cap}^{\prec}$
is continuous everywhere if $Y$ is a binary random variable.

\begin{figure}[t!]
\begin{centering}
\includegraphics[width=0.9\columnwidth]{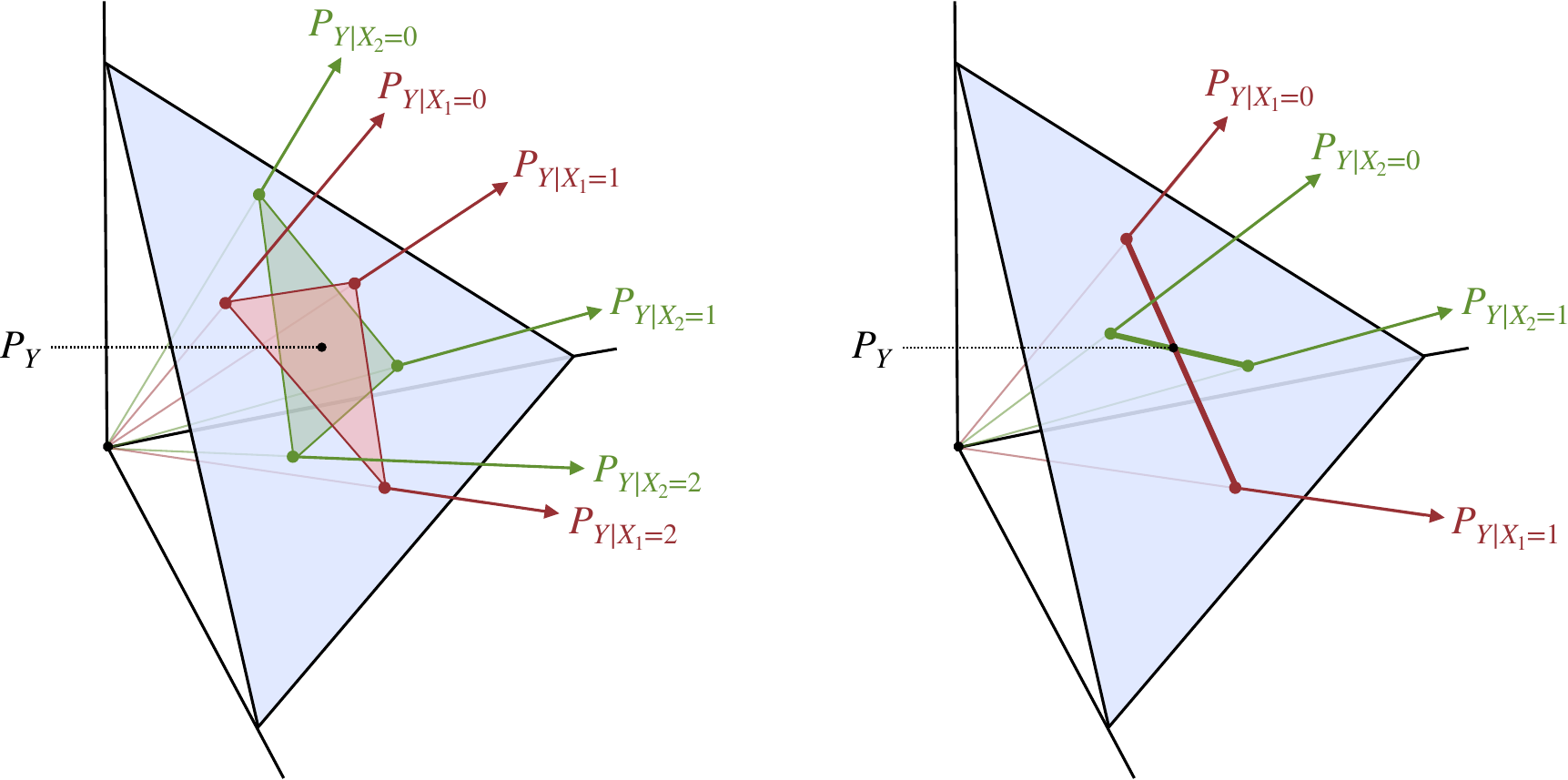}
\par\end{centering}
\caption{\label{fig:cont}Illustration of \cref{thm:cont}, which provides a
sufficient condition for the local continuity of $I_{\cap}^{\prec}$.
Consider two scenarios, both of which involves two sources $X_{1}$
and $X_{2}$ and a target $Y$ with cardinality $|\mathrm{\mathcal{Y}}|=3$.
The blue areas indicate the simplex of probability distributions over
$\mathrm{\mathcal{Y}}$, with the marginal $P_{Y}$ and the pairwise
conditionals $P_{Y\vert X_{i}=x_{i}}$ marked. On the left, both sources
have $\mathrm{rank\;}P_{Y\vert X_{i}}=3=|\mathrm{\mathcal{Y}}|$,
so $I_{\cap}^{\prec}$ is locally continuous. On the right, both sources
have $\mathrm{rank\;}P_{Y\vert X_{i}}=2<|\mathrm{\mathcal{Y}}|$,
so $I_{\cap}^{\prec}$ is not necessarily locally continuous. Note
that $I_{\cap}^{\prec}$ is also continuous if only source has $\mathrm{rank\;}P_{Y\vert X_{i}}=3$.}
\end{figure}

We illustrate the meaning of \cref{thm:cont} visually in \cref{fig:cont}
. We show two situations, both of which involve two sources $X_{1}$
and $X_{2}$ and a target $Y$ with cardinality $|\mathrm{\mathcal{Y}}|=3$.
In one situation, both pairwise conditional distributions have rank equal to $|\mathrm{\mathcal{Y}}|$, so $I_{\cap}^{\prec}$ is locally continuous. In the other situation,
both pairwise conditional distributions are rank deficient (e.g.,
this might happen because $X_{1}$ and $X_{2}$ have cardinality $|\mathcal{X}_{1}|=|\mathcal{X}_{2}|=2$),
so $I_{\cap}^{\prec}$ is not guaranteed to be continuous. From the
figure it is easy to see how the discontinuity may arise. Given the
definition of the Blackwell order and $I_{\cap}^{\prec}$, for any random variable $Q$ in the feasible set of \cref{eq:istarmaintext}, the conditional distributions $P_{Y\vert Q=q}$ 
must fall within the intersection of the distributions spanned by
$P_{Y\vert X_{1}}$ and $P_{Y\vert X_{2}}$ (the intersection of the
red and green shaded regions in \cref{fig:cont}). On the right, the
size of this intersection can discontinuously jump from a line (when
$P_{Y\vert X_{1}}$ and $P_{Y\vert X_{2}}$ are perfectly aligned)
to a point (when $P_{Y\vert X_{1}}$ and $P_{Y\vert X_{2}}$ are not
perfectly aligned). Thus, the discontinuity of $I_{\cap}^{\prec}$
arises from a geometric phenomenon, which is related to the discontinuity
of the intersection of low-dimensional vector subspaces.

We briefly comment on the continuity of $I_{\cup}^{\prec}$. As we
described above, this measure turns out to be equivalent to $I_{\cup}^{\mathrm{BROJA}}$
. The continuity of $I_{\cup}^{\mathrm{BROJA}}$ in the bivariate
case was proven in Theorem 35 of Ref.~\cite{banerjee_unique_2018}. We believe
that the continuity of $I_{\cup}^{\mathrm{BROJA}}$ for an arbitrary
number of sources can be shown using similar methods, although we
leave this for future work.

\subsection{Behavior on the COPY Gate}

\label{sec:identity}

As mentioned in \cref{sec:PID}, the ``COPY gate'' example is often
used to test the behavior of different redundancy measures. The COPY
gate has two sources, $X_{1}$ and $X_{2}$, and a target $Y=(X_{1},X_{2})$
which is a copy of the joint outcome. It is expected that redundancy
should vanish if $X_{1}$ and $X_{2}$ are statistically independent,
as formalized by the \emph{Independent identity} property in \cref{eq:iiprop}.

Blackwell redundancy $I_{\cap}^{\prec}$ satisfies the \emph{Independent
identity}. In fact, we prove a more general result, which shows that
$I_{\cap}^{\prec}(X_{1},X_{2}\!\shortrightarrow\!(X_{1},X_{2}))$
is equal to an information-theoretic measure called \emph{Gács-Körner
common information} $C(X\wedge Y)$ \citep{gacs_common_1973,wolf_zero-error_2004,griffith2014intersection}.
$C(X\wedge Y)$ quantifies the amount of information that can be deterministically
extracted from both random variables $X$ or $Y$, and it is 
closely related to the ``deterministic function'' order $\vartriangleleft$
defined in \cref{eq:funcrel-1}. Formally, it can be written as 
\begin{equation}
C(X\wedge Y)=\sup_{Q}H(Q)\qquad\text{such that}\qquad Q\vartriangleleft X,Q\vartriangleleft Y,\label{eq:comminfo}
\end{equation}
where $H$ is Shannon entropy. In \cref{app:misc}, we prove the following
result.
\begin{thm}
$I_{\cap}^{\prec}(X_{1},X_{2}\!\shortrightarrow\!(X_{1},X_{2}))=C(X_{1}\wedge X_{2})$.\label{thm:gacs}
\end{thm}

Note that $0\le C(X_{1}\wedge X_{2})\le I(X_{1};X_{2})$ \citep{gacs_common_1973},
so $I_{\cap}^{\prec}$ satisfies the \emph{Independent identity} property.
At the same time, $C(X_{1}\wedge X_{2})$ can be strictly less than
$I(X_{1};X_{2})$. For example, if $P_{X_{1}X_{2}}$ has full support,
then $I(X_{1};X_{2})$ can be arbitrarily large while $C(X_{1}\wedge X_{2})=0$
(see proof of \cref{thm:gacs}). This means that $I_{\cap}^{\prec}$
violates a previously proposed property, sometimes called the \emph{Identity
 property}, that suggests that redundancy should satisfy $I_{\cap}(X_{1};X_{2}\!\shortrightarrow\!(X_{1},X_{2}))=I(X_{1};X_{2})$.
However, the validity of the \emph{Identity property} is not clear,
and several papers have argued against it \citep{rauh2014reconsidering,james2018unique}. 

The value of $C(X_{1}\wedge X_{2})$ depends on the precise pattern
of zeros in the joint distribution $P_{X_{1}X_{2}}$ and is therefore
not continuous. For instance, for the bivariate COPY gate, redundancy
can change discontinuously as one goes from the situation where $X_{1}=X_{2}$
(so that all information is redundant, $I_{\cap}^{\prec}=I(X_{1};X_{2})$)
to one where $X_{1}$ and $X_{2}$ are almost, but not entirely, identical.
This discontinuity can be understood in terms of \cref{thm:cont} and
\cref{fig:cont}: in the COPY gate, the cardinality of the target variable
$|\mathrm{\mathcal{Y}}|=|\mathcal{X}_{1}|\times|\mathcal{X}_{2}|$
is larger than the cardinality of the individual sources. In other
words, when the sources $X_{1}$ and $X_{2}$ are not perfectly correlated,
they provide information about different ``subspaces'' of the target
$(X_{1},X_{2})$, and so it is possible that very little (or none)
of their information is redundant.

At the same time, the Blackwell property, \cref{thm:blackwell}, implies
that
\begin{equation}
I_{\cap}^{\prec}(X_{1},X_{2}\!\shortrightarrow\!X_{1})=I(X_{1};X_{2})=I_{\cap}^{\prec}(X_{1},X_{2}\!\shortrightarrow\!X_{2})\label{eq:eqmi}
\end{equation}
In other words, the redundancy in $X_{1}$ and $X_{2}$, where either
one of the individual sources is taken as the target, is given by
the mutual information $I(X_{1};X_{2})$. This holds even though the
redundancy in the COPY gate can be much lower than $I(X_{1};X_{2})$.

It is also interesting to consider how Blackwell union information,
$I_{\cup}^{\prec}$, behaves on the COPY gate. Using techniques from
\citep{bertschinger2014quantifying}, it can be shown that the union
information is simply the joint entropy,
\begin{align}
I_{\cup}^{\prec}(X_{1};X_{2}\!\shortrightarrow\!(X_{1},X_{2})) & =H(X_{1},X_{2}).\label{eq:unionmi}
\end{align}
Since $H(X_{1},X_{2})=I(X_{1},X_{2};X_{1},X_{2})$, \cref{eq:unionmi}
and \cref{eq:synDefM} together imply that the COPY gate has no synergy.

Note that we can use \cref{thm:gacs} and \cref{eq:unionmi} to illustrate
that $I_{\cap}^{\prec}$ and $I_{\cup}^{\prec}$ violate the inclusion-exclusion
principle, \cref{eq:iep-1}. Using \cref{eq:unionmi} and a bit of rearranging,
\cref{eq:iep-1} becomes equivalent to $I_{\cap}^{\prec}(X_{1};X_{2}\!\shortrightarrow\!(X_{1},X_{2}))\stackrel{?}{=}I(X_{1};X_{2})$,
which is the \emph{Identity property} mentioned above. $I_{\cap}^{\prec}$
violates this property, since redundancy for the COPY gate can be
smaller than $I(X_{1};X_{2})$.

\section{Examples and Comparisons to Previous Measures}

\label{sec:Relation-to-existing}




In this section, we compare our proposed measure of Blackwell redundancy
$I_{\cap}^{\prec}$ to existing redundancy measures. We focus on redundancy,
rather than union information, because redundancy has seen much more
development in the literature, and because Blackwell union information
$I_{\cup}^{\prec}$ is equivalent to an existing measure (see \cref{subsec:Blackwell-union-information-priorwork}).

\subsection{Qualitative Comparison}

In Table~\ref{tab:Comparison}, we compare $I_{\cap}^{\prec}$
to six existing measures of multivariate redundancy:
\begin{itemize}
\item $I_{\cap}^{\mathrm{WB}}$, the redundancy measure first proposed by
Williams and Beer \citep{williams2010nonnegative}.
\item $I_{\cap}^{\text{\ensuremath{\mathrm{MMI}}}}$, the ``minimum mutual
information'' \citep{barrett_exploration_2015}, \cref{eq:mmi} in
\cref{subsec:Relation-to-prior}.
\item $I_{\cap}^{\vartriangleleft}$, proposed by Griffith et al. \citep{griffith2014intersection},
\cref{eq:IwedgeDef} in \cref{subsec:Relation-to-prior}.
\item $I_{\cap}^{\mathrm{GH}}$, proposed by Griffith and Ho \citep{griffith_quantifying_2015},
\cref{eq:igh} in \cref{subsec:Relation-to-prior}.
\item $I_{\cap}^{\text{\ensuremath{\mathrm{Ince}}}}$, proposed by Ince
\citep{ince_measuring_2017}.
\item $I_{\cap}^{\mathrm{FL}}$, proposed by Finn and Lizier \citep{finn_pointwise_2018}.
\end{itemize}
We also compare $I_{\cap}^{\prec}$ to three existing measures of
bivariate redundancy (i.e., for 2 sources):
\begin{itemize}
\item $I_{\cap}^{\mathrm{BROJA}}$, proposed by Bertschinger et al.~\citep{bertschinger2014quantifying}, defined in \cref{eq:broja1}.
\item $I_{\cap}^{\text{\ensuremath{\mathrm{Harder}}}}$, proposed by Harder
et al. \citep{harder_bivariate_2013}.
\item $I_{\cap}^{\text{\ensuremath{\mathrm{dep}}}}$, proposed by James
et al. \citep{james2018unique}. 
\end{itemize}
For $I_{\cap}^{\prec}$ as well as the 9 existing measures, we consider
the following properties, which are chosen to highlight differences
between our approach and previous proposals:
\begin{enumerate}
\item Has it been defined for more than 2 sources
\item Does it obey the \emph{Monotonicity} axiom from \cref{subsec:Axiomatic-derivation}
\item Is it compatible with the inclusion-exclusion principle (IEP) for
the bivariate case, such that union information as defined in \cref{eq:iep-1}
obeys $I_{\cup}(X_{1};X_{2}\!\shortrightarrow\!Y)\le I(X_{1},X_{2};Y)$
\item Does it obey the \emph{Independent identity} property, \cref{eq:iiprop}
\item Does it obey the \emph{Blackwell} \emph{property} (possibly in its
multivariate form, \cref{thm:blackwell})
\end{enumerate}
We also consider two additional properties, which require a bit of
introduction. 

The first property was suggested by Ref.~\citep{bertschinger2014quantifying},
who argued that redundancy should only depend on the pairwise marginal
distributions of each source with the target,
\begin{equation}
\text{If }p_{X_{i}Y}=p_{\tilde{X}_{i}\tilde{Y}}\text{ for all \ensuremath{i}, then }I_{\cap}(X_{1};\dots;X_{n}\!\shortrightarrow\!Y)=I_{\cap}(\tilde{X}_{1};\dots;\tilde{X}_{n}\!\shortrightarrow\!\tilde{Y}).\label{eq:pairwiseProperty}
\end{equation}
In Table~\ref{tab:Comparison}, we term this property \emph{Pairwise
marginals.} We believe that the validity of \cref{eq:pairwiseProperty}
is not universal, but may depend on the particular setting in which
the PID is being used. However, redundancy redundancy measures that
satisfy this property have one important advantage: they are well-defined
not only when the sources are random variables $X_{1},\dots,X_{n}$,
but also in the more general case when the sources are channels $\kappa_{X_{1}|Y},\dots,\kappa_{X_{n}\vert Y}$.

\newcommand\colwidth{0.42cm}
\newcommand\colwidthh{.85cm}
\begin{table}[H]
\tablesize{\fontsize{9}{9}\selectfont}
\caption{Comparison of different redundancy measures.
?  indicate properties that we could not easily establish.}\label{tab:Comparison}

\begin{tabularx}{13cm}{c>{\centering}p{\colwidth}>{\centering}p{\colwidth}>{\centering}p{\colwidth}>{\centering}p{\colwidth}>{\centering}p{\colwidth}>{\centering}p{\colwidth}>{\centering}p{\colwidth}>{\centering}p{\colwidthh}>{\centering}p{\colwidthh}>{\centering}p{\colwidthh}}
\toprule 
 & $\boldsymbol{I_{\cap}^{\prec}}$ & $\boldsymbol{I_{\cap}^{\mathrm{WB}}}$ & $\boldsymbol{I_{\cap}^{\text{\ensuremath{\mathrm{MMI}}}}}$ & $\boldsymbol{I_{\cap}^{\vartriangleleft}}$ & $\boldsymbol{I_{\cap}^{\mathrm{GH}}}$ & $\boldsymbol{I_{\cap}^{\text{\ensuremath{\mathrm{Ince}}}}}$ & $\boldsymbol{I_{\cap}^{\mathrm{FL}}}$ & $\boldsymbol{I_{\cap}^{\mathrm{BROJA}}}$ & $\boldsymbol{I_{\cap}^{\text{\ensuremath{\mathrm{Harder}}}}}$ & $\boldsymbol{I_{\cap}^{\text{\ensuremath{\mathrm{dep}}}}}$\tabularnewline
\midrule 
 
More than 2 sources & $\checked$ & $\checked$ & $\checked$ & $\checked$ & $\checked$ & $\checked$ & $\checked$ &  &  & \tabularnewline
\midrule 
Monotonicity & $\checked$ & $\checked$ & $\checked$ & $\checked$ & $\checked$ &  &  & $\checked$ & $\checked$ & $\checked$\tabularnewline
\midrule 
IEP for bivariate case &  & $\checked$ & $\checked$ &  &  & ? & ? & $\checked$ & $\checked$ & $\checked$\tabularnewline
\midrule 
Independent identity & $\checked$ &  &  & $\checked$ & $\checked$ & $\checked$ &  & $\checked$ & $\checked$ & $\checked$\tabularnewline
\midrule 
Blackwell property & $\checked$ &  &  &  &  &  &  & $\checked$ & $\checked$ & \tabularnewline
\midrule 
Pairwise marginals & $\checked$ & $\checked$ & $\checked$ & $\checked$ &  &  &  & $\checked$ & $\checked$ & \tabularnewline
\midrule 
Target equality & $\checked$ & $\checked$ & $\checked$ &  & $\checked$ &  &  & $\checked$ & $\checked$ & \tabularnewline
\bottomrule 
\end{tabularx}%

\end{table}

The second property has not been previously considered
in the literature, although it appears to be highly intuitive. Observe
that the target random variable $Y$ contains all possible information
about itself. Thus, it may be expected that adding the target to the
set of sources should not decrease the redundancy:
\begin{equation}
I_{\cap}(X_{1};\dots;X_{n};Y\!\shortrightarrow\!Y)=I_{\cap}(X_{1};\dots;X_{n}\!\shortrightarrow\!Y).\label{eq:targetequality}
\end{equation}
In Table~\ref{tab:Comparison}, we term this property \emph{Target
equality}. Note that for redundancy measures which can be put in the form of
\cref{eq:IcapGen}, \emph{Target Equality }is satisfied if the order
$\sqsubset$ obeys $X_{i}\sqsubset Y$ for all sources $X_{i}$. (Note
also that \emph{Target Equality} is unrelated to the previously proposed
\emph{Strong Symmetry} property; for instance,
it is easy to show that the redundancy measures $I_{\cap}^{\mathrm{WB}}$
and $I_{\cap}^{\text{\ensuremath{\mathrm{MMI}}}}$ satisfy \emph{Target
Equality}, even though they violate \emph{Strong Symmetry} \citep{bertschinger2013shared}.) 

\subsection{Quantitative Comparison}

We now illustrate our proposed measure of redundancy $I_{\cap}^{\prec}$
on some simple examples, and compare its behavior to existing redundancy
measures. 

The values of $I_{\cap}^{\prec}$ were computed with our code, provided
at \citep{githubRepo}. The values of all other redundancy measures
except $I_{\cap}^{\mathrm{GH}}$ were computed using the
\texttt{dit} Python package \citep{dit}. To our knowledge, there
have been no previous proposals for how to compute $I_{\cap}^{\mathrm{GH}}$.
In fact, this measure involves maximizing a convex function subject
to linear constraints, and can be computed using similar methods as
$I_{\cap}^{\prec}$. We provide code for computing $I_{\cap}^{\mathrm{GH}}$
at \citep{githubRepo}.

We begin by considering some simple bivariate examples. In all cases,
the sources $X_{1}$ and $X_{2}$ are binary and uniformly distributed.
The results are shown in Table~\ref{tab:bivariateex}.

%
\renewcommand\colwidth{1.05cm}
\renewcommand\colwidthh{1.15cm}

\begin{table}[H]
\tablesize{\fontsize{8}{8}\selectfont}
\caption{\label{tab:bivariateex}Behavior of $I_{\cap}^{\prec}$ and other
redundancy measures on bivariate examples.}
\begin{adjustwidth}{-1.5cm}{0cm}

\begin{tabularx}{16cm}{c>{\centering}p{\colwidth}>{\centering}p{\colwidthh}>{\centering}p{\colwidth}>{\centering}p{\colwidth}>{\centering}p{\colwidth}>{\centering}p{\colwidth}>{\centering}p{\colwidth}>{\centering}p{\colwidth}>{\centering}p{\colwidth}>{\centering}p{\colwidth}}

\toprule 
\textbf{Target} & $\boldsymbol{I_{\cap}^{\prec}}$ & $\boldsymbol{I_{\cap}^{\mathrm{WB}}}$ & $\boldsymbol{I_{\cap}^{\mathrm{MMI}}}$ & $\boldsymbol{I_{\cap}^{\wedge}}$ & $\boldsymbol{I_{\cap}^{\mathrm{GH}}}$ & $\boldsymbol{I_{\cap}^{\text{\ensuremath{\mathrm{Ince}}}}}$ & $\boldsymbol{I_{\cap}^{\text{FL}}}$ & $\boldsymbol{\substack{I_{\cap}^{\mathrm{BROJA}}\\
I_{\cap}^{\text{\ensuremath{\mathrm{Harder}}}}}
}
$ & $\boldsymbol{I_{\cap}^{\text{\ensuremath{\mathrm{dep}}}}}$\tabularnewline
 
\midrule 
$Y=X_{1}\,\mathrm{AND}\,X_{2}$ & 0.311 & 0.311 & 0.311 & 0 & 0.123 & 0.104 & 0.561 & 0.311 & 0.082\tabularnewline
\midrule 
$Y=X_{1}+X_{2}$ & 0.5 & 0.5 & 0.5 & 0 & 0 & 0 & 0.5 & 0.5 & 0.189\tabularnewline
\midrule 
$Y=X_{1}$ & $I(X_{1};\!X_{2})$ & $I(X_{1};\!X_{2})$ & $I(X_{1};\!X_{2})$ & \mbox{$C(X_{1}\!\wedge\!X_{2})$} & $I(X_{1};\!X_{2})$ & {*} & 1 & $I(X_{1};\!X_{2})$ & $I(X_{1};\!X_{2})$\tabularnewline
\midrule 
$Y=(X_{1},X_{2})$ & \mbox{$C(X_{1}\!\wedge\!X_{2})$} & 1 & 1 & \mbox{$C(X_{1}\!\wedge\!X_{2})$} & \mbox{$C(X_{1}\!\wedge\!X_{2})$} & {*} & 1 & $I(X_{1};\!X_{2})$ & $I(X_{1};\!X_{2})$\tabularnewline
\bottomrule 
\end{tabularx}
\end{adjustwidth} 
\end{table}

\begin{enumerate}
\item The AND gate, $Y=X_{1}\mathrm{\text{ AND }}X_{2}$, with $X_{1}$
and $X_{2}$ independent. (It is incorrectly stated in Refs.~\citep{griffith_quantifying_2015,banerjee2015synergy}
that $I_{\cap}^{\mathrm{GH}}$ vanishes here; actually
$I_{\cap}^{\mathrm{GH}}(X_{1};X_{2}\!\shortrightarrow\!X_{1}\text{ AND }X_{2})\approx0.123$,
which corresponds to the maximum achieved in \cref{eq:igh} by $Q=X_{1}\text{ OR }X_{2}$.)
\item The SUM gate: $Y=X_{1}+X_{2}$, with $X_{1}$ and $X_{2}$ independent.
\item The UNQ gate: $Y=X_{1}$.  Here $I_{\cap}^{\text{\ensuremath{\mathrm{Ince}}}}$
(marked with $*$) gave values that increased with the amount of correlation
between $X_{1}$ and $X_{2}$ but were typically larger than $I(X_{1};\!X_{2})$.
\item The COPY gate: $Y=(X_{1},X_{2})$. Here, our redundancy measure
is equal to the Gács-Körner common information between $X$ and $Y$,
as discussed in \cref{sec:identity}. The same holds for the redundancy
measures $I_{\cap}^{\mathrm{GH}}$ and $I_{\cap}^{\vartriangleleft}$, which can be shown using
a slight modification of the proof of \cref{thm:gacs}. For this gate,
$I_{\cap}^{\text{\ensuremath{\mathrm{Ince}}}}$ (marked with $*$)
gave the same values as for the UNQ gate, which increased with the
amount of correlation between $X_{1}$ and $X_{2}$ but were typically
larger than $I(X_{1};X_{2})$.
\end{enumerate}
\label{sec:Examples}

We also analyze several examples with three sources, with the results
shown in Table~\ref{tab:trivariateex}. We considered those previously
proposed measures which can be applied to more than two sources (we
do not show $I_{\cap}^{\mathrm{GH}}$, as our implementation was too
slow for these examples).
\begin{enumerate}
\item Three-way AND gate: $Y=X_{1}\textrm{ AND }X_{2}\textrm{ AND }X_{3}$,
where the sources are binary and uniformly and independently distributed. 
\item Three-way SUM gate: $Y=X_{1}+X_{2}+X_{3}$, where the sources are
binary and uniformly and independently distributed. 
\item ``Overlap'' gate: we defined four independent uniformly distributed
binary random variables, $A,B,C,D$. These were grouped into three
sources $X_{1},X_{2},X_{3}$ as $X_{1}=(A,B)$, $X_{2}=(A,C)$, $X_{3}=(A,D)$.
The target was the joint outcome of all three sources, $Y=(X_{1},X_{2},X_{3})=((A,B),(A,C),(A,D))$.
Note that the three sources overlap on a single random variable
$A$, which suggests that the redundancy should be 1 bit.
\end{enumerate}

\begin{table}[H]
\caption{\label{tab:trivariateex}Behavior of $I_{\cap}^{\prec}$ and other
redundancy measures on three sources.}
\begin{tabularx}{\textwidth}{cCCCCCC}
\toprule 
\textbf{Target} & $\boldsymbol{I_{\cap}^{\prec}}$ & $\boldsymbol{I_{\cap}^{\mathrm{WB}}}$ & $\boldsymbol{I_{\cap}^{\mathrm{MMI}}}$ & $\boldsymbol{I_{\cap}^{\wedge}}$ & $\boldsymbol{I_{\cap}^{\text{\ensuremath{\mathrm{Ince}}}}}$ & $\boldsymbol{I_{\cap}^{\text{FL}}}$\tabularnewline
 
\midrule 
$Y=X_{1}\text{ AND }X_{2}\text{ AND }X_{3}$ & 0.138 & 0.138 & 0.138 & 0 & 0.024 & 0.294\tabularnewline
\midrule 
$Y=X_{1}+X_{2}+X_{3}$ & 0.311 & 0.311 & 0.311 & 0 & 0 & 0.561\tabularnewline
\midrule 
$Y=((A,B),(A,C),(A,D))$ & 1 & 2 & 2 & 1 & 1 & 2\tabularnewline
\bottomrule 
\end{tabularx}

\end{table}

\section{Discussion and Future Work}

\label{sec:Discussion}

In this paper, we proposed a new general framework for defining the
partial information decomposition (PID). Our framework was motivated
in several ways, including a formal analogy with intersections and unions
in set theory as well as an axiomatic derivation. 

We also used our general framework to propose concrete measures of
redundancy and union information, which have clear operational interpretations
based on Blackwell's theorem. Other PID measures, such as synergy
and unique information, can be computed from our measures of redundancy
and union information via simple expressions.

One unusual aspect of our framework is that it provides separate measures
of redundancy and union information. As we discuss above, most prior
work on the PID assumed that redundancy and union information are
related to each other via the so-called ``inclusion-exclusion''
principle. We argue that the inclusion-exclusion principle should
not be expected to hold in the context of the PID, and in fact that
it leads to counterintuitive behavior once 3 or more sources are
present. This suggests that different information decompositions should
be derived for redundancy vs. union information. This idea is related
to a recent proposal in the literature, which argues that two different
PIDs are needed, one based on redundancy and one based on synergy
\citep{chicharro_redundancy_2016}. An interesting direction for future
work is to relate our framework with the dual decompositions proposed
in \citep{chicharro_redundancy_2016}.

From a practical standpoint, an important direction for future work
is to develop better schemes for computing our redundancy measure.
This measure is defined in terms of a convex maximization problem,
which in principle can be NP-hard (a similar convex maximization problem
was proven to be NP-hard in \citep{kovacevic_entropy_2015}). Our
current implementation, which enumerates the vertices of the feasible
set, works well for relatively small state spaces, but we do not expect
it to scale to situations with many sources, or where the sources
have large cardinalities. However, the problem of convex maximization
with linear constraints is a very active area of optimization research,
with many proposed algorithms \citep{horst1984global,pardalos1986methods,benson1995concave}.
Investigating these algorithms, as well as various approximation schemes
such as relaxations and variational bounds, is of interest.

Finally, we showed how our framework can be used to define measures
of redundancy and union information in situations that go beyond the
standard setting of the PID (e.g., when the probability distribution
of the target is not specified). Our framework can even be applied
in domains beyond Shannon information theory, such as algorithmic
information theory and quantum information theory. Future work may
exploit this flexibility to explore various new applications
of the PID.
\vspace{6pt}

\acknowledgments{We thank Paul Williams, 
Alexander Gates, Nihat Ay, Bernat Corominas-Murtra, 
Pradeep Banerjee, and especially Johannes Rauh 
for helpful discussions and suggestions. 
We also thank the Santa Fe Institute for helping 
to support this research.}
\appendixtitles{yes}
\appendixstart{}

\appendix

\section{PID Axioms\label{app:axioms}}

In developing the PID framework, Williams and Beer \citep{williams2010nonnegative,williams_information_2011}
proposed that any measure of redundancy should obey a set of axioms.
In slightly modified form, these axioms can be written as follows:
\begin{itemize}
\item \emph{Symmetry}: $I_{\cap}(X_{1};\dots;X_{n}\!\shortrightarrow\!Y)$
is invariant to the permutation of $X_{1},\dots,X_{n}$.
\item \emph{Self-redundancy}: $I_{\cap}(X_{1}\!\shortrightarrow\!Y)=I(Y;X_{1})$.
\item \emph{Monotonicity}: $I_{\cap}(X_{1};\dots;X_{n}\!\shortrightarrow\!Y)\le I_{\cap}(X_{1};\dots;X_{n-1}\!\shortrightarrow\!Y)$.
\item \emph{Deterministic equality:} $I_{\cap}(X_{1};\dots;X_{n}\!\shortrightarrow\!Y)=I_{\cap}(X_{1};\dots;X_{n-1}\!\shortrightarrow\!Y)$
if $X_{i}=f(X_{n})$ for some $i<n$ and deterministic function $f$.
\end{itemize}
These axioms are based on intuitions regarding the behavior of intersection
in set theory \citep{williams_information_2011}. The \emph{Symmetry}
axiom is self-explanatory. \emph{Self-redundancy} states that if only
a single-source is present, all of its information is redundant. \emph{Monotonicity}
states that redundancy should not increase when an additional source
is considered (consider that the size of set intersection can only
decrease as more sets are considered). \emph{Deterministic equality
}states that redundancy should remain the same when an additional
source $X_{n}$ is added that contains all (or more) of the same information
that is already contained in an existing source $X_{i}$ (which is
formalized as the condition $X_{i}=f(X_{n})$).

Union information was considered the original PID proposal \citep{williams2011generalized,williams_information_2011},
as well as a more recent paper \citep{griffith2014quantifying}. \citep{griffith2014quantifying}
proposed that any measure of union information should satisfy the
following set of natural axioms, stated here in slightly modified
form:
\begin{itemize}
\item \emph{Symmetry}: $I_{\cup}(X_{1};\dots;X_{n}\!\shortrightarrow\!Y)$
is invariant to the permutation of $X_{1},\dots,X_{n}$.
\item \emph{Self-union}: $I_{\cup}(X_{1}\!\shortrightarrow\!Y)=I(Y;X_{1})$.
\item \emph{Monotonicity}: $I_{\cup}(X_{1};\dots;X_{n}\!\shortrightarrow\!Y)\ge I_{\cup}(X_{1};\dots;X_{n-1}\!\shortrightarrow\!Y)$.
\item \emph{Deterministic equality}: $I_{\cup}(X_{1};\dots;X_{n}\!\shortrightarrow\!Y)=I_{\cup}(X_{1};\dots;X_{n-1}\!\shortrightarrow\!Y)$
if $X_{n}=f(X_{i})$ for some $i<n$ and deterministic function $f$.
\end{itemize}
These axioms are based on intuitions concerning the behavior of the
union operator in set theory, and are the natural ``duals'' of the
redundancy axioms mentioned above.

\section{Uniqueness Proofs}

\label{app:unqproofs}

\begin{proof}[Proof of \cref{thm:unq}.]
Assume there is a redundancy measure $I_{\cap}'$ that obeys the
five axioms stated in the theorem. We will show that $I_{\cap}'=I_{\cap}$,
as defined in \cref{eq:IcapGen}.

Given \cref{eq:IcapGen} and the definition of the supremum, for any
$\epsilon>0$ there exists a random variable $Q$ such that $Q\sqsubset X_{i}$
for $i\in\{1,\dots,n\}$ and
\begin{equation}
I(Q;Y)\ge I_{\cap}(X_{1};\dots;X_{n}\!\shortrightarrow\!Y)-\epsilon,\label{eq:dasf2}
\end{equation}
By \emph{Order equality}, $I_{\cap}'(Q;X_{1};\dots;X_{k}\!\shortrightarrow\!Y)=I_{\cap}'(Q;X_{1};\dots;X_{k-1}\!\shortrightarrow\!Y)$.
Induction gives
\[
I_{\cap}'(Q;X_{1};\dots;X_{n}\!\shortrightarrow\!Y)=I_{\cap}'(Q\!\shortrightarrow\!Y)=I(Q;Y)\ge I_{\cap}(X_{1};\dots;X_{n}\!\shortrightarrow\!Y)-\epsilon
\]
where we used \emph{Self-redundancy} and \cref{eq:dasf2}. We also
have $I_{\cap}'(Q;X_{1};\dots;X_{n}\!\shortrightarrow\!Y)\le I_{\cap}'(X_{1};\dots;X_{n}\!\shortrightarrow\!Y)$
by \emph{Symmetry} and \emph{Monotonicity}. Combining gives
\[
I_{\cap}(X_{1};\dots;X_{n}\!\shortrightarrow\!Y)-\epsilon\le I_{\cap}'(X_{1};\dots;X_{n}\!\shortrightarrow\!Y).
\]
We now show that $I_{\cap}$ is the largest measure that satisfies
\emph{Existence}. Let $Q$ be a random variable that obeys $Q\sqsubset X_{i}$
for all $i\in\{1,\dots,n\}$ and $I_{\cap}'(X_{1};\dots;X_{n}\!\shortrightarrow\!Y)=I(Y;Q)$.
Since $Q$ falls within the feasible set of the optimization problem
in \cref{eq:IcapGen}, 
\[
I_{\cap}'(X_{1};\dots;X_{n}\!\shortrightarrow\!Y)=I(Q;Y)\le I_{\cap}(X_{1};\dots;X_{n}\!\shortrightarrow\!Y).
\]
Combining gives 
\[
I_{\cap}'(X_{1};\dots;X_{n}\!\shortrightarrow\!Y)-\epsilon\le I_{\cap}(X_{1};\dots;X_{n}\!\shortrightarrow\!Y)-\epsilon\le I_{\cap}'(X_{1};\dots;X_{n}\!\shortrightarrow\!Y).
\]
Since this holds for all $\epsilon>0$, taking the limit $\epsilon\to0$
gives $I_{\cap}'=I_{\cap}$.
\end{proof}

\begin{proof}[Proof of \cref{thm:unq-union}]
Assume there is a union information measure $I_{\cup}'$ that obeys
the five axioms stated in the theorem. We will show that $I_{\cup}'=I_{\cup}$,
as defined in \cref{eq:IcupGen}. 

Given \cref{eq:IcupGen} and the definition of the infinitum, for any
$\epsilon>0$ there exists a random variable $Q$ such that $Q\sqsubset X_{i}$
for $i\in\{1,\dots,n\}$ and 
\begin{equation}
I(Q;Y)\le I_{\cup}(X_{1};\dots;X_{n}\!\shortrightarrow\!Y)+\epsilon,\label{eq:dasf2-1}
\end{equation}
By \emph{Order equality}, $I_{\cup}'(Q;X_{1};\dots;X_{k}\!\shortrightarrow\!Y)=I_{\cup}'(Q;X_{1};\dots;X_{k-1}\!\shortrightarrow\!Y)$.
Induction gives 
\[
I_{\cup}'(Q;X_{1};\dots;X_{n}\!\shortrightarrow\!Y)=I_{\cup}'(Q\!\shortrightarrow\!Y)=I(Q;Y)\le I_{\cup}(X_{1};\dots;X_{n}\!\shortrightarrow\!Y)+\epsilon
\]
where we used \emph{Self-union} and \cref{eq:dasf2-1}. We also have
$I_{\cup}'(Q;X_{1};\dots;X_{n}\!\shortrightarrow\!Y)\ge I_{\cup}'(X_{1};\dots;X_{n}\!\shortrightarrow\!Y)$
by \emph{Symmetry} and \emph{Monotonicity}. Combining gives
\[
I_{\cup}(X_{1};\dots;X_{n}\!\shortrightarrow\!Y)+\epsilon\ge I_{\cup}'(X_{1};\dots;X_{n}\!\shortrightarrow\!Y).
\]
We now show that $I_{\cap}$ is the smallest measure that satisfies
\emph{Existence}. Let $Q$ be a random variable that obeys $X_{i}\sqsubset Q$
for all $i\in\{1,\dots,n\}$ and $I_{\cup}'(X_{1};\dots;X_{n}\!\shortrightarrow\!Y)=I(Y;Q)$.
Since $Q$ falls within the feasible set of the optimization problem
in \cref{eq:IcupGen},
\[
I_{\cup}'(X_{1};\dots;X_{n}\!\shortrightarrow\!Y)=I(Q;Y)\ge I_{\cup}(X_{1};\dots;X_{n}\!\shortrightarrow\!Y).
\]
Combining gives 
\[
I_{\cup}'(X_{1};\dots;X_{n}\!\shortrightarrow\!Y)\le I_{\cup}(X_{1};\dots;X_{n}\!\shortrightarrow\!Y)+\epsilon\le I_{\cap}'(X_{1};\dots;X_{n}\!\shortrightarrow\!Y)+\epsilon.
\]
Since this holds for all $\epsilon>0$, taking the limit $\epsilon\to0$
gives $I_{\cup}'=I_{\cup}$.
\end{proof}

\section{Computing $I_{\cap}^{\prec}$}

\label{app:opt}

Here we consider the optimization problem that defines our proposed
measure of redundancy, \cref{eq:istarmaintext}. We first prove a bound
on the required cardinality of $Q$.
\begin{thm}
\label{thm:cardinality}For optimizing \cref{eq:istarmaintext}, it
suffices to consider $Q$ with cardinality $\left|\mathrm{\mathcal{Q}}\right|=\left(\sum_{i}\left|\mathcal{X}_{i}\right|\right)-n+1$.
\end{thm}

\begin{proof}
Consider any random variable $Q$ with outcome set $\mathrm{\mathcal{Q}}$
which satisfies $Q\prec_{Y}X_{i}$ for all $i$. We show that whenever
$Q$ has full support on $\left|\mathrm{\mathcal{Q}}\right|>\left(\sum_{i}\left|\mathcal{X}_{i}\right|\right)-n+1$
outcomes, there is another random variable $\tilde{Q}$ which achieves $I(\tilde{Q};Y)\ge I(Q;Y)$, while
satisfying 
$\tilde{Q}\prec_{Y}X_{i}$ for all $i$ and having support on at most $\left(\sum_{i}\left|\mathcal{X}_{i}\right|\right)-n+1$
outcomes.

To begin, let $\Omega$ indicate the set of random variables over
outcomes $\mathrm{\mathcal{Q}}$, such that all $\tilde{Q}\in\Omega$
satisfy:
\begin{align}
P_{Y\vert\tilde{Q}}(y\vert q) & =P_{Y\vert Q}(y\vert q)\qquad &  & \text{for all \ensuremath{y,q\in\{q\in\mathrm{\mathcal{Q}}:P_{\tilde{Q}}(q)>0\}}}\label{eq:condY}\\
\sum_{q}P_{\tilde{Q}}(q)P_{X_{i}\vert Q}(x_{i}\vert q) & =P_{X_{i}}(x_{i})\qquad &  & \text{for all }i,x_{i}.\label{eq:condM}
\end{align}
Since $Q\prec_{Y}X_{i}$, by \cref{eq:garbl} there exist channels
$\kappa_{Q\vert X_{i}}(q\vert x_{i})$ that satisfy $P_{Q\vert Y}(q\vert y)=\sum_{x_{i}}\kappa_{Q\vert X_{i}}(q\vert x_{i})P_{X_{i}|Y}(x_{i}\vert y).$
Now write the conditional distribution over $\tilde{Q}$ and $Y$
as 
\begin{align}
P_{\tilde{Q}Y}(q,y)=\frac{P_{\tilde{Q}}(q)}{P_{Q}(q)}P_{QY}(q,y) & =\sum_{x_{i}}\frac{P_{\tilde{Q}}(q)}{P_{Q}(q)}\kappa_{Q\vert X_{i}}(q\vert x_{i})P_{X_{i}|Y}(x_{i}\vert y)P_{Y}(y)\nonumber \\
 & =\sum_{x_{i}}\kappa_{\tilde{Q}\vert X_{i}}'(q\vert x_{i})P_{X_{i}|Y}(x_{i}\vert y)P_{Y}(y),\label{eq:garbapp}
\end{align}
where we used \cref{eq:condY} and defined the channel $\kappa_{\tilde{Q}\vert X_{i}}'$
as 
\[
\kappa_{\tilde{Q}\vert X_{i}}'=\frac{P_{\tilde{Q}}(q)}{P_{X_{i}}(x_{i})}\left[\frac{P_{X_{i}}(x_{i})}{P_{Q}(q)}\kappa_{Q\vert X_{i}}(q\vert x_{i})\right],
\]
(Note this is a kind of double Bayesian inverse, given \cref{eq:condM}.)
\cref{eq:garbapp} implies that $\tilde{Q}\prec_{Y}X_{i}$ for all
$i$.

We now show that there is $\tilde{Q}\in\Omega$ that achieves $I(Q;Y)\le I(\tilde{Q};Y)$
and has support on at most $\left(\sum_{i}\left|\mathcal{X}_{i}\right|\right)-n+1$
outcomes in $\mathrm{\mathcal{Q}}$. Write the mutual information
between any $\tilde{Q}\in\Omega$ and $Y$ as
\begin{equation}
I(\tilde{Q};Y)=\sum_{q}P_{\tilde{Q}}(q)D_{\mathrm{KL}}(P_{Y\vert\tilde{Q}=q}\Vert P_{Y})=\sum_{q}P_{\tilde{Q}}(q)D_{\mathrm{KL}}(P_{Y\vert Q=q}\Vert P_{Y}),\label{eq:klin}
\end{equation}
where $D_{\mathrm{KL}}$ is the Kullback-Leibler divergence. We consider
the maximum of this mutual information across $\Omega$, $I^{*}=\max_{\tilde{Q}\in\Omega}I(\tilde{Q};Y)$.
Using \cref{eq:condM} and \cref{eq:klin}, this maximum can be written
as 
\begin{align*}
I^{*} & =\max_{\omega\in\Delta}\sum_{q}\omega(q)D(P_{Y\vert Q=q}\Vert P_{Y})\quad\text{such that}\quad\forall i,x_{i}:\sum_{q}\omega(q)P_{X_{i}\vert Q}(x_{i}\vert q)=P_{X_{i}}(x_{i}),
\end{align*}
where $\Delta$ is the set of all distributions over $\mathrm{\mathcal{Q}}$.
By conservation of probability, $\sum_{x_{i}}P_{X_{i}}(x_{i})=1$,
so we can eliminate a constraint for one of the outcomes $x_{i}$
of each source $i$. Thus, $I^{*}$ is the maximum of a linear function
over $\Delta$, subject to $\sum_{i}(\left|\mathcal{X}_{i}\right|-1)=\left(\sum_{i}\left|\mathcal{X}_{i}\right|\right)-n$
hyperplane constraints. 

The feasible set is compact, and the maximum will be achieved at one
of the extreme points of the feasible set. By Dubin's Theorem \citep{dubins1962extreme},
any extreme point of this feasible set can be expressed as a convex
combination of at most $\left(\sum_{i}\left|\mathcal{X}_{i}\right|\right)-n+1$
extreme points of $\Delta$. In other words, the maximum in \cref{eq:istarmaintext}
is achieved by a random variable $\tilde{Q}$ with support on at most
$\left(\sum_{i}\left|\mathcal{X}_{i}\right|\right)-n+1$ values of
$\mathrm{\mathcal{Q}}$. This random variable satisfies 
\[
I(\tilde{Q};Y)=I^{*}\ge I(Q;Y),
\]
where the last inequality comes from the fact that $Q$ is an element
of $\Omega$.
\end{proof}
We now return to the optimization problem in \cref{eq:istarmaintext}.
Given \cref{thm:cardinality} and the definition of the Blackwell order
in \cref{eq:garbl}, it can be rewritten as
\begin{align}
I_{\cap}^{\prec}(X_{1};\dots;X_{n}\!\shortrightarrow\!Y)= & \max_{\kappa_{Q\vert Y},\kappa_{Q\vert X_{1},\dots,}\kappa_{Q\vert X_{n}}}I_{\kappa}(Q,Y)\label{eq:istarapp}\\
 & \text{such that}\;\forall i,y,x_{i}\,:\sum_{x_{i}}\kappa_{Q\vert X_{i}}(q\vert x_{i})P_{X_{i}\vert Y}(x_{i}\vert y)=\kappa_{Q\vert Y}(q\vert y).\nonumber 
\end{align}
where the optimization is over channels with $\mathrm{\mathcal{Q}}$
of cardinality $\left(\sum_{i}\left|\mathcal{X}_{i}\right|\right)-n+1$.
The notation $I_{\kappa}(Q;Y)$ indicates the mutual information that
arises from the marginal distribution $P_{Y}$ and the conditional
distribution $\kappa_{Q\vert Y}$,
\[
I_{\kappa}(Q;Y)=\sum_{y}P_{Y}(y)\kappa_{Q\vert Y}(q\vert y)\ln\frac{\kappa_{Q\vert Y}(q\vert y)}{\sum_{y'}\kappa_{Q\vert Y}(q\vert y')P_{Y}(y')}
\]
\cref{eq:istarapp} involves maximizing a convex function over the
convex polytope defined by the following system of linear (in)equalities:
\begin{align}
 &  & \Lambda=\Big\{(\kappa_{Q\vert Y},\kappa_{Q\vert X_{1},\dots,}\kappa_{Q\vert X_{n}}):\nonumber \\
 &  & \forall i,x_{i},q\quad & \kappa_{Q\vert X_{i}}(q\vert x_{i})\ge0,\label{eq:fes-1}\\
 &  & \forall q,y\quad & \kappa_{Q\vert Y}(q\vert y)\ge0,\label{eq:fes-2}\\
 &  & \forall y\quad & \sum_{q}\kappa_{Q\vert Y}(q\vert y)=1,\label{eq:fes3-1}\\
 &  & \forall i,x_{i}\quad & \sum_{q}\kappa_{Q\vert X_{i}}(q\vert x_{i})=1,\label{eq:fes4-1}\\
 &  & \forall i,y,q\in\mathrm{\mathcal{Q}}\setminus\{0\}\quad & \left[\sum_{x_{i}}\kappa_{Q\vert X_{i}}(q\vert x_{i})P_{X_{i}Y}(x_{i},y)\right]-\kappa_{Q\vert Y}(q\vert y)P_{Y}(y)=0\Big\},\label{eq:fes5-1}
\end{align}
We do not place a constraint on $q=0$ in \cref{eq:fes5-1} because
that would be redundant with the constraints \cref{eq:fes3-1} and
\cref{eq:fes4-1}. Also, note that we replaced the $\sup$ in \cref{eq:istarmaintext}
with $\max$ in \cref{eq:istarapp}, which is justified since we are
optimizing over a finite dimensional, closed, and bounded region (thus
the supremum is always achieved).

The maximum of a convex function over a convex polytope is found at
one of the vertices of the polytope. To find the solution to
\cref{eq:istarapp}, we use a computational geometry package to enumerate
the vertices of $\Lambda$. We evaluate $I_{\kappa}(Y;Q)$ at each
vertex, and pick the maximum value. This procedure also finds optimal
conditional distributions $\kappa_{Q\vert Y},\kappa_{Q\vert X_{1},\dots,}\kappa_{Q\vert X_{n}}$.
Code is available at \citep{githubRepo}.

\section{Continuity of $I_{\cap}^{\prec}$\label{app:continuity}}

To prove the continuity of $I_{\cap}^{\prec}$, we begin by considering
the feasible set of the optimization problem in \cref{eq:istarapp},
as specified by the system (in)equalities in \cref{eq:fes-1,eq:fes-2,eq:fes3-1,eq:fes4-1,eq:fes5-1}.
For convenience, write this system of (in)equalities in matrix notation,
\begin{equation}
\Lambda=\left\{ \vec{\kappa}\in\mathbb{R}^{|\mathrm{\mathcal{Q}}||\mathrm{\mathcal{Y}}|+\sum_{i}|\mathrm{\mathcal{Q}}||\mathcal{X}_{i}|}:\kappa\ge0,A\kappa=a\right\} ,\label{eq:fmatrix-1}
\end{equation}
where $\vec{\kappa}$ is a vector representation of $(\kappa_{Q\vert Y},\kappa_{Q\vert X_{1},\dots,}\kappa_{Q\vert X_{n}})$,
\textls[-5]{the matrix $A$ encodes the left-hand side of \cref{eq:fes3-1,eq:fes4-1,eq:fes5-1},
and the vector $a$ is filled with 1s and 0s, as appropriate.} 

We first prove the following lemma.
\begin{lem}
\label{lem:contlem}The matrix $A$ defined in \cref{eq:fmatrix-1}
is full rank if $n-1$ or more of the pairwise conditional distributions
have $\mathrm{rank\;}P_{Y\vert X_{i}}=|\mathrm{\mathcal{Y}}|$.
\end{lem}

\begin{proof}
Without loss of generality, assume that $P_{Y}$ has full support
(otherwise none of the pairwise marginals $P_{X_{i}Y}$ can achieve
rank $|\mathrm{\mathcal{Y}}|$). Write $A$ in block matrix form as
$A=\left[\begin{array}{c}
B\\
C
\end{array}\right]$, where the matrix $B$ has $|\mathrm{\mathcal{Y}}|+\sum_{i}|\mathcal{X}_{i}|$
rows and encodes the constraints of \cref{eq:fes3-1} and \cref{eq:fes4-1},
and the matrix $C$ has $n|\mathrm{\mathcal{Y}}|(|\mathrm{\mathcal{Q}}|-1)$
rows and encodes the constraints of \cref{eq:fes5-1}.

Each row in $B$ has a 1 in some column which is zero in every other
row of $B$ and every row of $C$. This column corresponds either
to $\kappa_{Q\vert Y}(0\vert y)$ for a particular $y$ (for constraints
like \cref{eq:fes3-1}), or to $\kappa_{Q\vert X_{i}}(0\vert x_{i})$
for a particular $i$ and $x_{i}$ (for constraints like \mbox{\cref{eq:fes4-1}}).
These columns are 0 in $C$ because $q=0$ is omitted \cref{eq:fes5-1}.
This means that no row of $B$ is a linear combination of other rows
in $B$ or $C$, and that no row in $C$ is a linear combination of
any set of other rows that includes a row in $B$. Therefore, if the
rows of $A$ are linearly dependent, it must be that the rows of $C$
are linearly dependent.

Next, let $\vec{c}^{i,y,q}$ indicate the row of $C$ that represents
the constraints in \cref{eq:fes5-1} for some source $i$ and outcomes
$y,q\ne0$. Any such row has a column for each $x_{i}\in\mathcal{X}_{i}$
with value $P_{X_{i}Y}(x_{i},y)$ (at the same index as the row in
$\vec{\kappa}$ that represents $\kappa_{Q\vert X_{i}}(q\vert x_{i})$).
Since $P_{X_{i}Y}(x_{i},y)>0$ for at least one $x_{i}\in\mathcal{X}_{i}$,
one of these columns must be non-zero. At the same time, these columns
are zero in every row $\vec{c}^{j,y,q'}$ where $j\ne i$ or $q'\ne q$.
This means that row $\vec{c}^{i,y,q}$ can only be a linear combination
of other rows in $C$ if, for all $x_{i}$, $P_{X_{i}Y}(x_{i},y)$
is a linear combination of $P_{X_{i}Y}(x_{i},y')$ for $y'\ne y$.
In linear algebra terms, this can be stated as $\mathrm{rank\;}P_{Y\vert X_{i}}<|\mathrm{\mathcal{Y}}|$.

The previous argument shows that if $A$ is linearly dependent, there
must be at least one source $i$ with $\mathrm{rank\;}P_{Y\vert X_{i}}<|\mathrm{\mathcal{Y}}|$
and some row $\vec{c}^{i,y,q}$ which is a linear combination of other
rows from $C$. Observe that this row $\vec{c}^{i,y,q}$ has a column
with value $P_{Y}(y)>0$ (at the same index as the row in $\vec{\kappa}$
that represents $\kappa_{Q\vert Y}(q\vert y)$). This column is zero
in every other row $\vec{c}^{i,y',q'}$ for $y'\ne y$ or $q'\ne q$.
This means that $\vec{c}^{i,y,q}$ is a linear combination of a set
of other rows in $C$ that include some row $\vec{c}^{j,y,q}$ for
$j\ne i$. This implies that $\vec{c}^{j,y,q}$ is also a linear combination
of other rows in $C$, which means that $\mathrm{rank\;}P_{Y\vert X_{j}}<|\mathrm{\mathcal{Y}}|$.

We have shown that if $A$ is linearly dependent, there must be at
least two pairwise conditionals with $\mathrm{rank\;}P_{Y\vert X_{i}}<|\mathrm{\mathcal{Y}}|$.
\end{proof}
We are now ready to prove \cref{thm:cont}.
\begin{proof}[Proof of \cref{thm:cont}]
For the case of a single source ($n=1$), $I_{\cap}^{\prec}$ reduces
to the mutual information $I_{\cap}^{\prec}=I(Y;X_{1})$, which is
continuous (Section 2.3, \cite{yeung2012first}). Thus, without loss
of generality, we assume that $n\ge2$.

Next, we define some notation. Note that the optimum value ($I_{\cap}^{\prec}$)
and the feasible set of the optimization problem in \cref{eq:istarapp}
is a function of the pairwise marginal distributions $P_{X_{1}Y},\dots,P_{X_{n}Y}$.
We write $\Omega$ for the set of all pairwise marginal distributions
which have the same marginal over $Y$: 
\[
\Omega=\Big\{(q_{X_{1}Y},\dots,q_{X_{n}Y}):\sum_{x_{i}}q_{X_{i}Y}(x_{i},y)=\sum_{x_{j}}q_{X_{j}Y}(x_{j},y)\quad\forall i,j\Big\}.
\]
For any $r\in\Omega$, let $I_{\cap}^{\prec}(r)$ indicate the corresponding
optimum value in \cref{eq:istarapp}, given the marginals in $r$,
and let $\Lambda(r)$ indicate the feasible set of the optimization
problem, as defined in \cref{eq:fmatrix-1}. 

Note that the matrix $A$ in \cref{eq:fmatrix-1} depends on the choice
of $r$, which we indicate by writing it as the matrix-valued function
$A(r)$. Given any $r=(q_{X_{1}Y},\dots,q_{X_{n}Y})\in\Omega$ and
feasible solution $\kappa=(\kappa_{Q\vert Y},\kappa_{Q\vert X_{1},\dots,}\kappa_{Q\vert X_{n}})\in\Lambda(r)$,
let $I(r,\kappa)$ indicate the corresponding mutual information $I(Q;Y)$,
where the marginal distribution over $Y$ is specified by $r$ and
the conditional distribution of $Q$ given $Y$ is specified by $\kappa_{Q\vert Y}$.
Using this notation, $I_{\cap}^{\prec}(r)=\max_{\kappa\in\Lambda(r)}I(r,\kappa)$. 

Below, we show that $I_{\cap}^{\prec}(r)$ is continuous if $r$ is
\emph{rank regular}~\citep{lewis2009semicontinuity}, which means
that there is a neighborhood $U\subseteq\Omega$ of $r$ such that
$\mathrm{rank\;}A(r')=\mathrm{rank\;}A(r)$ for all $r'\in U$. Then,
to prove the theorem, we assume that $A(r)$ is full rank. Given \cref{lem:contlem},
this is true as long as $n-1$ or more of the pairwise conditionals
$P_{Y\vert X_{i}}$ have $\mathrm{rank\;}P_{Y\vert X_{i}}=|\mathrm{\mathcal{Y}}|$.
Note that a matrix $M$ is full rank iff the singular values $\sigma(M)$
are all strictly positive. Since $A(r)$ is full rank, and $A(r)$
and $\sigma(M)$ are continuous, there is a neighborhood $U$ of $r$
such that the singular values $\sigma(A(r'))$ are all strictly positive
for all $r'\in U$, therefore all $A(r'))$ have full rank. This shows
that $r$ is rank regular and so $I_{\cap}^{\prec}$ is continuous
at $r$.

We now prove that $I_{\cap}^{\prec}(r)$ is continuous if $A(r)$
is rank regular. To do so, we will use Hoffman's Theorem \citep{hoffman1952approximate,daniel_perturbations_1973}.
In our case, it states that for any pair of marginals $r,r'\in\Omega$
and a feasible solution $\kappa'\in\Lambda(r')$, there exists a feasible
solution $\kappa\in\Lambda(r)$ such that
\begin{equation}
\left\Vert \kappa-\kappa'\right\Vert \le\alpha\left\Vert A(r)-A(r')\right\Vert ,\label{eq:hoffman}
\end{equation}
where $\alpha$ is a constant that does not depend on $r'$ or $\kappa'$.
(In the notation of \citep{daniel_perturbations_1973}, we take $G=G'$,
$g=g'$ and $d'=d$, and use that the norm of $s=\kappa'$ is bounded,
given that it is finite dimensional and has entries in $[0,1]$).
We will also use Daniel's theorem \citep[Thm. 4.2, ][]{daniel_perturbations_1973},
which states that for any $r,r'\in\Omega$ such that $\mathrm{rank\;}A(r)=\mathrm{rank\;}A(r')$,
and any feasible solution $\kappa\in\Lambda(r)$, there exists $\kappa'\in\Lambda(r')$
such that
\begin{equation}
\left\Vert \kappa-\kappa'\right\Vert \le\beta\left\Vert A(r)-A(r')\right\Vert ,\label{eq:daniels}
\end{equation}
where $\beta$ is a constant that doesn't depend on $r'$ (in the
notation of \citep{daniel_perturbations_1973}, $\varepsilon'=\left\Vert A(r)-A(r_{i}')\right\Vert $
and again use that $\kappa$ have a bounded norm). 

Now consider also any sequence $r_{1}',r_{2}',\dots\in\Omega$ that
converges to a marginal $r\in\Omega$. Let $\kappa_{i}'\in\Lambda(r_{i}')$
indicate an optimal solution of \cref{eq:istarapp} for $r_{i}$, so
that $I(r_{i}',\kappa_{i}')=I_{\cap}^{\prec}(r_{i}')$. Given \cref{eq:hoffman},
there is a corresponding sequence $\kappa_{1},\kappa_{2},\dots\in\Lambda(r)$
such that 
\[
\left\Vert \kappa_{i}-\kappa_{i}'\right\Vert \le\alpha\left\Vert A(r)-A(r_{i}')\right\Vert .
\]
Since $A(\cdot)$ is continuous and $r_{i}'$ converges to $r$, we
have $\lim_{i\to\infty}A(r_{i}')=A(r)$ and therefore $\lim_{i\to\infty}\left\Vert \kappa_{i}-\kappa_{i}'\right\Vert =0$.
This implies
\begin{equation}
0=\lim_{i\to\infty}I(r_{i}',\kappa_{i}')-I(r,\kappa_{i})\ge\lim_{i\to\infty}I_{\cap}^{\prec}(r_{i}')-I_{\cap}^{\prec}(r)\label{eq:lbound1}
\end{equation}
where we first used continuity of mutual information, $I(r_{i}',\kappa_{i}')=I_{\cap}^{\prec}(r_{i}')$
and $I(r,\kappa_{i})\le I_{\cap}^{\prec}(r)$.

Now assume that $r$ is rank regular. Since $r_{i}$ converges to
$r$, $\mathrm{rank\;}A(r_{i}')=\mathrm{rank\;}A(r)$ for all sufficiently
large $i$. Let $\kappa\in\Lambda(r)$ be an optimal solution of \cref{eq:istarapp}
for $r$, so that $I(r,\kappa)=I_{\cap}^{\prec}(r)$. Given \cref{eq:daniels},
for all sufficiently large $i$ there exists $\kappa_{i}'\in\Lambda(r_{i}')$
such that
\[
\left\Vert \kappa-\kappa_{i}'\right\Vert \le\beta\left\Vert A(r)-A(r_{i}')\right\Vert .
\]
As before, we have $\lim_{i\to\infty}A(r_{i}')=A(r)$ and $\lim_{i\to\infty}\left\Vert \kappa-\kappa_{i}'\right\Vert =0$,
which implies
\begin{equation}
0=\lim_{i\to\infty}I(r_{i}',\kappa_{i}')-I(r,\kappa)\le\lim_{i\to\infty}I_{\cap}^{\prec}(r_{i}')-I_{\cap}^{\prec}(r)\label{eq:ubound1}
\end{equation}
where we first used continuity of mutual information, $I(r_{i}',\kappa_{i}')\le I_{\cap}^{\prec}(r_{i}')$,
and $I(r,\kappa)\le I_{\cap}^{\prec}(r)$.

Combining \cref{eq:lbound1} and \cref{eq:ubound1} proves continuity,
$\lim_{i\to\infty}I_{\cap}^{\prec}(r_{i})=I_{\cap}^{\prec}(r)$, under
the assumption that $A(r)$ is rank regular.

Finally, note that $A(r)$ is a real analytic function of $r$. This
means that almost all $r$ rank regular, because those $r$ which
are not rank regular form a proper analytic subset of $\Omega$ (which
has measure zero) \citep{lewis2009semicontinuity}. Thus, $I_{\cap}^{\prec}(r)$
is continuous almost everywhere.
\end{proof}
We finish our analysis of the continuity of $I_{\cap}^{\prec}$ by
showing global continuity when the target is a binary random variable.
\begin{cor}
\label{cor:binarycont}$I_{\cap}^{\prec}(X_{1};\dots;X_{n}\!\shortrightarrow\!Y)$
is continuous everywhere when $Y$ is a binary random variable.
\end{cor}

\begin{proof}
In an overloading of notation, let $I_{\cap}^{\prec}(r)$ and $I_{r}(X_{i};Y)$
indicate $I_{\cap}^{\prec}(X_{1};\dots;X_{n}\!\shortrightarrow\!Y)$
and the mutual information $I(X_{i};Y)$, respectively, for the joint
distribution $r_{X_{1}\dots X_{n}Y}$. By \cref{thm:cont}, $I_{\cap}^{\prec}$
can only be discontinuous at the joint distribution $P_{X_{1}\dots X_{n}Y}$
if there is a source $X_{i}$ with $\mathrm{rank\;}P_{Y\vert X_{i}}=1<|\mathrm{\mathcal{Y}}|$.
However, if source $X_{i}$ has rank 1, then the conditional distributions
$P_{Y\vert X_{i}=x_{i}}$ are the same for all $x_{i}$, so $I_{P}(X_{i};Y)=0$
and $I_{\cap}^{\prec}(P)=0$ (since $0\le I_{\cap}^{\prec}(P)\le I_{P}(X_{i};Y)$).
Finally, consider any sequence of joint distributions $s_{X_{1}\dots X_{n}Y}^{(n)}$
for $n=1,2,\dots$ that converges to $P_{X_{1}\dots X_{n}Y}$. We
have 
\[
0\le\lim_{n\to\infty}I_{\cap}^{\prec}(s^{(n)})\le\lim_{n\to\infty}I_{s^{(n)}}(X_{i};Y)=I_{P}(X_{i};Y)=0,
\]
where we used the continuity of mutual information. This shows that
$\lim_{n\to\infty}I_{\cap}^{\prec}(s^{(n)})=0=I_{\cap}^{\prec}(P)$,
proving continuity. 
\end{proof}

\section{Behavior of $I_{\cap}^{\prec}$ on Gaussian Random Variables}

\label{app:gaussian}

Although in this paper we focused on random variables with finite
sets of outcomes, we can briefly comment on the behavior of Blackwell
redundancy on Gaussian random variables. Suppose that all sources
$X_{1},\dots,X_{n}$ and the target $Y$ are continuous-valued, and
that the pairwise marginals $P_{X_{i}Y}$ are multivariate Gaussians.
In addition, suppose that $Y$ is one-dimensional (the sources $X_{i}$
can be multi-dimensional). Given these assumptions, Barrett \citep{barrett_exploration_2015}
analyzed the $I_{\cup}^{\mathrm{BROJA}}$ measure and showed that
the corresponding excluded information obeys $E(X_{j}\!\shortrightarrow\!Y\vert X_{i};X_{j})=0$
whenever $I(X_{i};Y)\le I(X_{j};Y)$. Recall that $I_{\cup}^{\mathrm{BROJA}}$
is equivalent to Blackwell union information $I_{\cup}^{\prec}$.
Then, given the Blackwell property, \cref{thm:blackwell-union}, and
the data processing inequality, \cref{eq:dpi}, the result in Ref.~\citep{barrett_exploration_2015}
implies that $X_{i}\prec_{Y}X_{j}$ if and only if $I(X_{i};Y)\le I(X_{j};Y)$.
Thus, for Gaussian random variables, Blackwell redundancy $I_{\cap}^{\prec}$
is equivalent to $I_{\cap}^{\text{\ensuremath{\mathrm{MMI}}}}$ redundancy,
as defined in \cref{eq:mmi}. This parallels the case for most other
redundancy measures \citep{barrett_exploration_2015}.

\section{Operational Interpretation of the $I_{\cap}^{\mathrm{GH}}$}

\label{app:Operational-interpretation-of-GH}

As mentioned in the main text, the redundancy measure $I_{\cap}^{\mathrm{GH}}$
is a special case of \cref{eq:IcapGen}, where the ``more informative''
order $B\sqsubset C$ is defined in terms of conditional independence
$B-C-Y$. Here we show that this ordering relation can be given an
operational interpretation, which is similar but distinct from the
operational interpretation of the Blackwell order $\prec_{Y}$ discussed
in \cref{subsec:The-Blackwell-order}.

To introduce this operational interpretation, let the random variable
$Y$ represent the state of the environment, and assume there are
two random variables $B$ and $C$ which have some information about
$Y$. Suppose that an agent tries to maximize expected utility $u(a,y)$
by using a strategy that depends either on the outcomes of $B$ or
$C$. Blackwell's theorem tells us that $B\prec_{Y}C$ iff an agent
with access to $C$ can always achieve higher expected utility than
an agent with access to $B$. It is possible, however, the agent with
access to $C$ may do worse than the agent with access to $B$, \emph{conditional
on the event that random variable $C$ has some particular outcome
$c$}. In the following theorem, we show $B-C-Y$ iff the agent cannot
do better with $B$ than $C$, even when conditioned on any particular
outcome $C=c$. (We thank Johannes Rauh for suggesting this simplified
proof.)

\begin{thm}
\label{thm:bl2}Given random variables $B$, $C$, and $Y$, $B-C-Y$
if and only if
\begin{equation}
\max_{\kappa_{A\vert B}}\sum_{y,a,b}P_{YB|C}(y,b\vert c)\kappa_{A\vert B}(a\vert b)u(a,y)\le\max_{\kappa_{A\vert C}}\sum_{y,a}P_{Y|C}(y\vert c)\kappa_{A\vert C}(a\vert c)u(a,y).\label{eq:max0}
\end{equation}
for all utility functions $u(a,y)$ and all $c\in\mathcal{C}$ with $P_C(c)>0$.
\end{thm}

\begin{proof}
Consider any $c\in\mathcal{C}$ with $P_C(c)>0$. By multiplying both sides of \cref{eq:max0} by $P_C(c)$ and rearranging, this inequality can be rewritten as 
\begin{align}
&\max_{\kappa_{A\vert B}}\sum_{y,a,b}P_{Y}(y)P_{C|Y}(c\vert y) P_{B|Y,C}(b\vert y,c)\kappa_{A\vert B}(a\vert b)u(a,y) \nonumber\\
&\qquad\qquad \le\max_{\kappa_{A\vert C}}\sum_{y,a}P_{Y}(y)P_{C|Y}(c\vert y)\kappa_{A\vert C}(a\vert c)u(a,y).
\label{eq:fdfg3}
\end{align}
Note that if \cref{eq:max0} holds for a given $c$ and any utility function $u(a,y)$, then it must also hold for that $c$ and any utility function of the form $u'(a,y):=P_{C|Y}(c\vert y)u(a,y)$. Plugging into  \cref{eq:fdfg3} gives
\begin{align}
\max_{\kappa_{A\vert B}}\sum_{y,a,b}P_{Y}(y) P_{B|Y,C}(b\vert y,c)\kappa_{A\vert B}(a\vert b)u'(a,y)\le\max_{\kappa_{A\vert C}}\sum_{y,a}P_{Y}(y)\kappa_{A\vert C}(a\vert c)u'(a,y).
\label{eq:fdfg4}
\end{align}
Now define two random variables: a constant
random variable $\hat{C}_{c}$ with a single outcome $c$ and $\hat{B}_{c}$
with the same outcomes as $B$ but having the conditional distribution
$P_{\hat{B}_{c}\vert Y}=P_{B\vert Y,C=c}$. Then, \cref{eq:fdfg4} can be written  in terms of these random variables as
\begingroup\makeatletter\def\f@size{9.5}\check@mathfonts
\def\maketag@@@#1{\hbox{\m@th\fontsize{10}{10}\selectfont\normalfont#1}}
\begin{equation}
\max_{\kappa_{A\vert B}}\sum_{y,a,b}P_{Y}(y)P_{\hat{B}_{c}|Y}(b\vert y)\kappa_{A\vert B}(a\vert b)u'(a,y)\le\max_{\kappa_{A\vert C}}\sum_{y,a}P_{Y}(y)P_{\hat{C}_{c}|Y}(c\vert y)\kappa_{A\vert C}(a\vert c)u'(a,y).\label{eq:max0-1}
\end{equation}
\endgroup
Given \cref{eq:eu,eq:blthm}, \cref{eq:max0-1} holds for all $u'$ iff $\hat{B}_{c}\prec_{Y}\hat{C}_c$.
Since $\hat{C}_c$ has a single outcome, it is independent of $Y$. That means $\hat{B}_{c}$ must be also independent of $Y$ and so  $P_{\hat{B}_c\vert Y=y} = P_{B\vert Y=y,C=c}$ is the same for all $y$, implying that $P_{B\vert Y=y,C=c}=P_{B\vert C=c}$. Since this holds for
all $c\in\mathcal{C}$,  $P_{B\vert YC}=P_{B\vert C}$ and 
therefore $B-C-Y$.
\end{proof}
Given \cref{thm:bl2}, $I_{\cap}^{\mathrm{GH}}$ can be given the following
operational interpretation.
Imagine two agents, Alice and Bob, who can acquire
information about $Y$ via different random variables, and then use
this information to maximize their expected utility. Suppose that
Alice has access to one of the sources $X_{i}$. Then,  $I_{\cap}^{\mathrm{GH}}$ is the maximum information that Bob
can have about $Y$ without being able to do better than Alice on
any utility function, regardless of which source $X_i$ Alice has access
to, and even when conditioned
on $X_{i}$ having any particular outcome $x_{i}$.

\section{Equivalence of $I_{\cup}^{\prec}$ and $I_{\cup}^{\mathrm{BROJA}}$}

\label{app:brojaequivalent}

The following proves that $I_{\cup}^{\prec}$ and $I_{\cup}^{\mathrm{BROJA}}$,
as defined via the optimization problems in \cref{eq:ustar,eq:brojaGenOpt},
are equivalent.
\begin{thm}
$I_{\cup}^{\prec}(X_{1};\dots;X_{n}\!\shortrightarrow\!Y)=I_{\cup}^{\mathrm{BROJA}}(X_{1};\dots;X_{n}\!\shortrightarrow\!Y).$
\end{thm}

\begin{proof}
Let $\tilde{X}_{1},\dots,\tilde{X}_{n}$ be a set of random variables
that achieve $I(Y;\tilde{X}_{1},\dots,\tilde{X}_{n})=I_{\cup}^{\mathrm{BROJA}}(X_{1};\dots;X_{n}\!\shortrightarrow\!Y)$.
Define the random variable $Q:=(\tilde{X}_{1},\dots,\tilde{X}_{n})$,
and note that $\tilde{X}_{i}\prec_{Y}Q$ for all $i$. Since $P_{\tilde{X}_{i}Y}=P_{X_{i}Y}$
for all $i$, it must be that $X_{i}\prec_{Y}Q$ for all $i$. Thus
$Q$ satisfies the constraints of the optimization problem in \cref{eq:ustar},
so
\begin{equation}
I_{\cup}^{\prec}(X_{1};\dots;X_{n}\!\shortrightarrow\!Y)\le I(Y;Q)=I_{\cup}^{\mathrm{BROJA}}(X_{1};\dots;X_{n}\!\shortrightarrow\!Y).\label{eq:lbound8}
\end{equation}
Next, consider the optimization in \cref{eq:ustar}. For any $\epsilon>0$,
let $Q$ be a random variable that satisfies $X_{i}\prec_{Y}Q$ and
achieves 
\begin{equation}
I(Y;Q)\le I_{\cup}^{\prec}(X_{1};\dots;X_{n}\!\shortrightarrow\!Y)+\epsilon.\label{eq:pra0}
\end{equation}
For each $i$, let $\kappa_{X_{i}\vert Q}$ be a channel that obeys
$P_{X_{i}\vert Y}(x_{i}\vert y)=\sum_{q}\kappa_{X_{i}\vert Q}(x_{i}\vert q)P_{Q\vert Y}(q\vert y)$
(such a channel must exist since $X_{i}\prec_{Y}Q$). Define the random
variables $\tilde{X}_{1},\dots,\tilde{X}_{n}$ with the joint distribution
\begin{equation}
P_{YQ\tilde{X}_{1}\dots\tilde{X}_{n}}(y,q,x_{1},\dots,x_{n})=P_{Y}(y)P_{Q\vert Y}(q\vert y)\prod_{i}\kappa_{X_{i}\vert Q}(x_{i}\vert q).\label{eq:dfadf2}
\end{equation}
Note that the pairwise marginals obey $P_{\tilde{X}_{i}Y}=P_{X_{i}Y}$.
Thus, all of the $\tilde{X}_{i}$ satisfy the marginal constraints in the right
hand side of \cref{eq:brojaGenOpt}, so
\begin{equation}
I_{\cup}^{\mathrm{BROJA}}(X_{1};\dots;X_{n}\!\shortrightarrow\!Y)\le I(Y;\tilde{X}_{1},\dots,\tilde{X}_{n}).\label{eq:prA1}
\end{equation}
By elementary properties of mutual information, we have
\begin{equation}
I(Y;\tilde{X}_{1},\dots,\tilde{X}_{n})\le I(Y;Q,\tilde{X}_{1},\dots,\tilde{X}_{n})\label{eq:prA2}
\end{equation}
Given \cref{eq:dfadf2}, the Markov condition $Y-Q-\tilde{X}_{1},\dots,\tilde{X}_{n}$
holds, so 
\begin{equation}
I(Y;\tilde{X}_{1},\dots,\tilde{X}_{n})\le I(Y;Q)\label{eq:prA3}
\end{equation}
by the data processing inequality. Combining \cref{eq:pra0,eq:prA1,eq:prA2,eq:prA3} implies
\[
I_{\cup}^{\mathrm{BROJA}}(X_{1};\dots;X_{n}\!\shortrightarrow\!Y)\le I(Y;Q)\le I_{\cup}^{\prec}(X_{1};\dots;X_{n}\!\shortrightarrow\!Y)+\epsilon.
\]
Since this holds for all $\epsilon$, we can take the limit $\epsilon\to0$
to give $I_{\cup}^{\mathrm{BROJA}}(X_{1};\dots;X_{n}\!\shortrightarrow\!Y)\le I_{\cup}^{\prec}(X_{1};\dots;X_{n}\!\shortrightarrow\!Y)$.
The result follows by combining with \cref{eq:lbound8}.
\end{proof}

\section{Relation between $I_{\cap}^{\mathrm{WB}}$ and Our General Framework}

\label{app:IwbGen}

Here we consider whether the redundancy measure $I_{\cap}^{\mathrm{WB}}$
proposed by Williams and Beer \citep{williams2010nonnegative} can
be put in the general form \cref{eq:IcapGen}. This measure is defined
as 
\begin{equation}
I_{\cap}^{\mathrm{WB}}(X_{1};\dots;X_{n}\!\shortrightarrow\!Y):=\sum_{y}P_{Y}(y)\min_{i}I(X_{i};Y=y),\label{eq:wbDef}
\end{equation}
where $I(X_{i};Y=y)$ is called the ``specific information'' between
$X_{i}$ and target outcome $y$,
\[
I(X_{i};Y=y):=D_{\mathrm{KL}}(P_{X_{i}\vert Y=y}\Vert P_{X_{i}})=\sum_{x_{i}}P_{X_{i}\vert Y}(x_{i}\vert y)\log\frac{P_{X_{i}\vert Y}(x_{i}\vert y)}{P_{X_{i}}(x_{i})},
\]
and $D_{\mathrm{KL}}$ is Kullback-Leibler (KL) divergence. 

Specific information obeys $I(X;Y)=\sum_{y}P_{Y}(y)I(X;Y=y)$.
Thus, \cref{eq:wbDef} looks similar to a mutual information expression,
where each specific information term is given by the smallest specific
information that $y$ carries about any of the sources. Motivated
by this interpretation, one might ask whether there exists a random variable $Q$
whose specific information terms are equal to $I(Q;Y=y)=\min_{i}I(X_{i};Y=y)$
for each $y$. If such a random variable existed, then $I_{\cap}^{\mathrm{WB}}$
could be written as
\begin{equation}
I_{\cap}^{\mathrm{WB}}(X_{1};\dots;X_{n}\!\shortrightarrow\!Y)\stackrel{?}{=}\max_{Q}I(Q;Y)\;\;\mathrm{\text{such that}}\;\forall i,y:I(Q;Y=y)\le I(X_{i};Y=y),\label{eq:iwbgen}
\end{equation}
which has the form of \cref{eq:IcapGen}, with the $\sqsubset$ order
defined as 
\begin{equation}
A\sqsubset B\;\text{iff\;}I(A;Y=y)\le I(B;Y=y)\;\;\text{for all \ensuremath{y\in\mathrm{\mathcal{Y}}}.}\label{eq:ppWB}
\end{equation}

Here we provide a counterexample to demonstrate that such a variable
does not exist in general, and so therefore \cref{eq:iwbgen} is not
generally valid. Suppose $Y$ has three outcomes $\mathrm{\mathcal{Y}}=\{0,1,2\}$
with a uniform distribution, and consider two binary sources $X_{1},X_{2}$
with the following conditional distributions,
\begin{align*}
P_{X_{1}\vert Y}(x_{1}\vert y) & =\begin{cases}
\delta(x_{1},y) & \text{if \ensuremath{y\in\{0,1\}}}\\
\frac{1}{2}\delta(x_{1},0)+\frac{1}{2}\delta(x_{1},1) & \text{if \ensuremath{y=2}}
\end{cases}\\
P_{X_{2}\vert Y}(x_{1}\vert y) & =\begin{cases}
\delta(x_{1},y) & \text{if \ensuremath{y\in\{0,2\}}}\\
\frac{1}{2}\delta(x_{1},0)+\frac{1}{2}\delta(x_{1},2) & \text{if \ensuremath{y=1}}
\end{cases}
\end{align*}
In this case, a simple calculation shows that the specific information
obeys (in bits)
\[
\begin{aligned}I(X_{1};Y=0)=1\qquad & I(X_{2};Y=0)=1\\
I(X_{1};Y=1)=1\qquad & I(X_{2};Y=1)=0\\
I(X_{1};Y=2)=0\qquad & I(X_{2};Y=2)=1
\end{aligned}
\]
Plugging into \cref{eq:wbDef} gives $I_{\cap}^{\mathrm{WB}}(X_{1};X_{2}\!\shortrightarrow\!Y)=1/3$.

Now consider the optimization problem in \cref{eq:iwbgen}. Since $I(X_{1};Y=2)=I(X_{2};Y=1)=0$,
any allowed $Q$ must satisfy $I(Q;Y=1)=I(Q;Y=2)=0$ and therefore
$P_{Q\vert Y=1}=P_{Q}=P_{Q\vert Y=2}$. Combined with the marginalization
identity $P_{Y}(0)P_{Q\vert Y=0}+P_{Y}(1)P_{Q\vert Y=1}+P_{Y}(2)P_{Q\vert Y=2}=P_{Q}$,
this implies that $P_{Q\vert Y=0}=P_{Q}$ and therefore that $I(Q;Y=0)=0$.
Thus, any allowed $Q$ obeys $I(Q;Y)=0\ne I_{\cap}^{\mathrm{WB}}$.
This means that $I_{\cap}^{\mathrm{WB}}$ cannot be expressed in the
form of \cref{eq:IcapGen} when $\sqsubset$ is defined as \cref{eq:ppWB}.

\section{Miscellaneous Derivations}

\label{app:misc}

\begin{proof}[Proof of \cref{lem:iep}]
We use a modified version of the example in \citep{rauh2014reconsidering,bertschinger2013shared}.
Consider a set of $n\ge3$ sources. The inclusion-exclusion principle
states that
\begin{equation}
I_{\cup}(X_{1};\dots;X_{n}\!\shortrightarrow\!Y)=\sum_{J\subseteq\{1,\dots,n\}\setminus\{\varnothing\}}(-1)^{\left|J\right|-1}I_{\cap}(X_{J_{1}};\dots;X_{J_{\left|J\right|}}\!\shortrightarrow\!Y).\label{eq:iepproof0}
\end{equation}
Now, let $X_{1},\dots,X_{n-1}$ be uniformly distributed and statistically
independent binary random variables, and take $X_{n}=X_{1}\;\mathrm{XOR}\;X_{2}$
and $Y=(X_{1},X_{2},X_{n})$. Note that $I(Y;X_{i})=1\text{ bit}$
for $i\in\{1,2,n\}$ and $I(Y;X_{i})=0$ for $i\in\{3,\dots,n-1\}$,
and that $I(Y;X_{1},\dots,X_{n})=2\text{ bit}$. Thus, $I_{\cap}(X_{i};X_{j}\!\shortrightarrow\!Y)=0$
whenever $i\in\{3,\dots,n-1\}$ or $j\in\{3,\dots,n-1\}$, as follows
from\emph{ Symmetry}, \emph{Self-redundancy}, and\emph{ Monotonicity}.
Note also that the outcomes of $Y$ are simply a relabelling of $(X_{1},X_{2})$,
and similarly for $(X_{1},X_{n})$ and $(X_{2},X_{n})$. Then, since
by \emph{Independent identity} property, $I_{\cap}(X_{i};X_{j}\!\shortrightarrow\!Y)=0$
for $i\ne j$ where $i,j\in\{1,2,n\}$. Thus, $I_{\cap}(X_{i};X_{j}\!\shortrightarrow\!Y)=0$
for all pairs $i\ne j$. By \emph{Monotonicity},
redundancy is 0 for any set of 2 or more sources.

Plugging this into \cref{eq:iepproof0} gives
\[
I_{\cup}(X_{1};\dots;X_{n}\!\shortrightarrow\!Y)=\sum_{i}I_{\cap}(X_{i}\!\shortrightarrow\!Y)=\sum_{i}I(X_{i}\!\shortrightarrow\!Y)=3\text{ bit}
\]
Note that this exceeds the total amount of information about the target
provided jointly by all sources, which is only $2$ bits, so $I_{\cup}(X_{1};\dots;X_{n}\!\shortrightarrow\!Y)\not\le I(Y;X_{1},\dots,X_{n})$.
\end{proof}

\begin{proof}[Proof of \cref{thm:blackwell}]
Without loss of generality, let $i=1$. We will use that $U^{\prec}(X_{1}\!\shortrightarrow\!Y\vert X_{1};\dots;X_{n})=0$
is equivalent to 
\begin{equation}
I_{\cap}^{\prec}(X_{1};\dots;X_{n}\!\shortrightarrow\!Y)=I(X_{1};Y).\label{eq:das3}
\end{equation}
We will use that by monotonicity of mutual information with respect
to $\prec_{Y}$ (see \cref{subsec:RedundancyDef}), 
\begin{equation}
I(X_{1};Y)\ge I_{\cap}^{\prec}(X_{1};\dots;X_{n}\!\shortrightarrow\!Y).\label{eq:df4}
\end{equation}

We first prove the ``if'' direction. Since $Q=X_{1}$ is in the
feasible set of \cref{eq:istarmaintext}, $I_{\cap}^{\prec}(X_{1};\dots;X_{n}\!\shortrightarrow\!Y)\ge I(X_{1};Y)$.
Combining with \cref{eq:df4} gives \cref{eq:das3}.

We now prove the ``only if'' direction. As described in \cref{app:opt},
$I_{\cap}^{\prec}$ can be expressed as an optimization over a finite
dimensional, closed, and bounded region, so the supremum in \cref{eq:istarmaintext}
is achieved. Thus, there is some $Q$ such that $Q\prec_{Y}X_{i}$
for all $i$ and
\begin{align}
I(Y;Q) & = I_{\cap}^{\prec}(X_{1};\dots;X_{n}\!\shortrightarrow\!Y).\label{eq:a0}
\end{align}
Since $Q\prec_{Y}X_{1}$, there is a conditional probability distribution
$\kappa_{\mathrm{\mathcal{Q}}\vert\mathcal{X}_{1}}$ such that $P_{Q\vert Y}(q\vert y)=\sum_{x_{1}}\kappa_{Q\vert X_{1}}(q\vert x_{1})P_{X_{1}\vert Y}(x_{1}\vert y)$.
Define a random variable $\tilde{Q}$ with the joint distribution
\[
P_{YX_{1}\tilde{Q}}(y,x_{1},q)=\kappa_{Q\vert X_{1}}(q\vert x_{1})P_{YX_{1}}(y,x_{1}).
\]
We will use that $P_{QY}=P_{\tilde{Q}Y}$. Then, the chain rule
for mutual information gives
\begin{align*}
I(Y;X_{1},\tilde{Q}) & =I(Y;\tilde{Q})+I(Y;X_{1}\vert\tilde{Q}) =I(Y;X_{1})+I(Y;\tilde{Q}\vert X_{1})=I(Y;X_{1}),
\end{align*}
where we used the Markov condition $Y-X_{1}-\tilde{Q}$.
Combining and rearranging gives
\begin{equation}
I(Y;\tilde{Q})=I(Y;X_{1})-I(Y;X_{1}\vert\tilde{Q}).\label{eq:a2}
\end{equation}
Now assume that \cref{eq:das3} holds. Combining with \cref{eq:a0}
and $P_{QY}=P_{\tilde{Q}Y}$ gives $I(Y;X_{1})=I(Y;Q)=I(Y;\tilde{Q})$.
Combining with \cref{eq:a2} gives $I(Y;X_{1}\vert\tilde{Q})=0$, meaning
that the Markov condition $Y-\tilde{Q}-X_{1}$ holds and therefore
$X_{1}\prec_{Y}\tilde{Q}$. Since 
$Q\prec_{Y}X_{i}$ for all $i$ and 
$P_{QY}=P_{\tilde{Q}Y}$, it also
the case that $\tilde{Q}\prec_{Y}X_{i}$ for all $i$. Finally, since  $\prec_{Y}$ is 
transitive, $X_{1}\prec_{Y} X_i$ for all $i$, which is the desired result.
\end{proof}

\begin{proof}[Proof of \cref{thm:blackwell-union}]
Without loss of generality, let $i=1$. We will use that $E^{\prec}(X_{1}\!\shortrightarrow\!Y\vert X_{1};\dots;X_{n})=0$
is equivalent to 
\begin{equation}
I_{\cup}^{\prec}(X_{1};\dots;X_{n}\!\shortrightarrow\!Y)=I(X_{1};Y).\label{eq:das4}
\end{equation}
We will use that by monotonicity of mutual information with respect
to $\prec_{Y}$ (see \cref{subsec:RedundancyDef}), 
\begin{equation}
I(X_{1};Y)\le I_{\cup}^{\prec}(X_{1};\dots;X_{n}\!\shortrightarrow\!Y).\label{eq:df4-1}
\end{equation}

We first prove the ``if'' direction. Since $Q=X_{1}$ is in the
feasible set of \cref{eq:ustar}, $I_{\cup}^{\prec}(X_{1};\dots;X_{n}\!\shortrightarrow\!Y)\le I(X_{1};Y)$.
Combining with \cref{eq:df4-1} gives \cref{eq:das4}.

We now prove the ``only if'' direction. As we show in \cref{app:brojaequivalent},
$I_{\cup}^{\prec}$ is equivalent to $I_{\cup}^{\mathrm{BROJA}}$,
which is defined as an optimization over a finite dimensional, closed,
and bounded region. Thus the infimum in \cref{eq:ustar} is always
achieved, so there is some $Q$ such that $X_{i}\prec_{Y}Q$ for all
$i$ and
\begin{align}
I(Y;Q) & =I_{\cup}^{\prec}(X_{1};\dots;X_{n}\!\shortrightarrow\!Y).\label{eq:a0-1}
\end{align}
Moreover, since $X_{1}\prec_{Y}Q$, there is a conditional probability
distribution $\kappa_{X_{1}\vert Q}$ such that $P_{X_{1}\vert Y}(x_{1}\vert y)=\sum_{q}\kappa_{X_{1}\vert Q}(x_{1}\vert q)P_{Q\vert Y}(q\vert y)$.
Define a random variable $\tilde{X}_{1}$ with the joint distribution
\[
P_{Y\tilde{X}_{1}Q}(y,x_{1},q)=\kappa_{X_{1}\vert Q}(x_{1}\vert q)P_{QY}(q,y).
\]
We will use that $P_{X_{1}Y}=P_{\tilde{X}_{1}Y}$. Then, using the
chain rule for mutual information, 
\begin{align*}
I(Y;\tilde{X}_{1},Q) & =I(Y;\tilde{X}_{1})+I(Y;Q\vert\tilde{X}_{1}) =I(Y;Q)+I(Y;\tilde{X}_{1}\vert Q)=I(Y;Q),
\end{align*}
where we used the Markov condition $Y-Q-\tilde{X}_{1}$.
Combining and rearranging gives
\begin{equation}
I(Y;\tilde{X}_{1})=I(Y;Q)-I(Y;Q\vert\tilde{X}_{1}).\label{eq:a2-2}
\end{equation}
Now assume that \cref{eq:das4} holds. Combining with \cref{eq:a0-1}
and $P_{X_{1}Y}=P_{\tilde{X}_{1}Y}$ gives $I(Y;X_{1})=I(Y;\tilde{X}_{1})=I(Y;Q)$.
Combining with \cref{eq:a2-2} gives $I(Y;Q\vert\tilde{X}_{1})=0$,
meaning that the Markov condition $Y-\tilde{X}_{1}-Q$ holds and therefore
$Q\prec_{Y}\tilde{X}_{1}$. Since $P_{X_{1}Y}=P_{\tilde{X}_{1}Y}$,
it is also the case that $Q\prec_{Y}X_{1}$. Finally, since 
$X_{i} \prec_{Y} Q$ for all $i$ and  $\prec_{Y}$ is 
transitive, $X_{i}\prec_{Y} X_1$ for all $i$, which is the desired result.
\end{proof}

\begin{proof}[Proof of \cref{thm:gacs}.]
Consider any random variable $Q$ which achieves the maximum in \cref{eq:istarmaintext}.
This implies there are channels $\kappa_{Q\vert X_{1}}$ and $\kappa_{Q\vert X_{2}}$
such that for any $q\in\mathrm{\mathcal{Q}}$ and $(x_{1},x_{2})\in\mathcal{X}_{1}\times\mathcal{X}_{2}$
with $p_{X_{1}X_{2}}(x_{1},x_{2})>0$,
\begin{align*}
P_{Q\vert X_{1}X_{2}}(q\vert x_{1},x_{2}) & =\sum_{x_{1}'}\kappa_{Q\vert X_{1}}(q\vert x_{1}')P_{X_{1}\vert X_{1}X_{2}}(x_{1}'\vert x_{1},x_{2})\\
P_{Q\vert X_{1}X_{2}}(q\vert x_{1},x_{2}) & =\sum_{x_{2}'}\kappa_{Q\vert X_{2}}(q\vert x_{2}')P_{X_{2}\vert X_{1}X_{2}}(x_{2}'\vert x_{1},x_{2}).
\end{align*}
We now equate the above two expressions, while using that $P_{X_{1}\vert X_{1}X_{2}}(x_{1}'\vert x_{1},x_{2})=\delta(x_{1}',x_{1})$
and $P_{X_{2}\vert X_{1}X_{2}}(x_{2}'\vert x_{1},x_{2})=\delta(x_{2}',x_{2})$
(where $\delta(\cdot,\cdot)$ is the Kronecker delta). This gives
\begin{equation}
\kappa_{Q\vert X_{1}}(q\vert x_{1})=\kappa_{Q\vert X_{2}}(q\vert x_{2})\label{eq:s1}
\end{equation}
for all $q$ and any $(x_{1},x_{2})$ where $p_{X_{1}X_{2}}(x_{1},x_{2})>0$.

Now consider a bipartite graph with vertex set $\mathcal{X}_{1}\cup\mathcal{X}_{2}$
and an edge between vertex $x_{1}$ and vertex $x_{2}$ if $P_{X_{1}X_{2}}(x_{1},x_{2})>0$.
Define $\Pi$ to be the set of connected components of this bipartite
graph, and let $f_{1}:\mathcal{X}_{1}\to\Pi$ be a function that maps
each $x_{1}$ to its corresponding connected component (for any $x_{1}$
with $P_{X_{1}}(x_{1})=0$, $f_{1}(x_{1})$ can be any value). \cref{eq:s1}
implies that if $x_{1}$ and $x_{1}'$ both belong to the same connected
component, then the constraint \cref{eq:s1} will ``propagate'' from
$x_{1}$ to $x_{1}'$, so that $\kappa_{Q\vert X_{1}}(q\vert x_{1})=\kappa_{Q\vert X_{1}}(q\vert x_{1}')$.
Said differently, this means that $\kappa_{Q\vert X_{1}}(q\vert x_{1})=\kappa_{Q\vert X_{1}}(q\vert f_{1}(x_{1}))$
and that the Markov condition $(X_{1},X_{2})-X_{1}-f_{1}(X_{1})-Q$
holds. This gives 
\begin{equation}
I(X_{1},X_{2};Q)\le I(X_{1},X_{2};f_{1}(X_{1}))=H(f_{1}(X_{1})),\label{eq:nd2}
\end{equation}
where the first inequality uses the data processing inequality, and
the second equality uses that $f_{1}(X_{1})$ is a deterministic function
of $(X_{1},X_{2})$. The upper bound in \cref{eq:nd2} is achieved
when $Q=f_{1}(X_{1})$, thus $Q\vartriangleleft X_{1}$. A similar argument shows that $Q\vartriangleleft X_{2}$. 

We have shown that the constraints in \cref{eq:istarmaintext} can be replaced by $Q\vartriangleleft X_{1}, Q\vartriangleleft X_{2}$. It is also clear that any $Q$ which is a deterministic function of either  $X_1$ or $X_2$ must also be a deterministic function of the target $Y=(X_1,X_2)$, hence $I(Y;Q)=H(Q)$. Combining these results shows that  \cref{eq:istarmaintext} is equivalent to 
\cref{eq:comminfo} for the COPY gate.
\end{proof}

\reftitle{References}
%
%

\end{document}